\documentclass[fleqn,usenatbib]{mnras}

\usepackage{newtxtext,newtxmath}
\usepackage[normalem]{ulem}
\usepackage[percent]{overpic}
\usepackage{array,multirow,graphicx}
\usepackage{comment}
\usepackage{booktabs}
\usepackage{tabularx,colortbl}

\newcommand{\be}{\begin{equation}}
\newcommand{\ee}{\end{equation}}
\newcommand{\bea}{\begin{eqnarray}}
\newcommand{\eea}{\end{eqnarray}}

\newcommand{\qmarks}[1]{``#1''}

\usepackage[T1]{fontenc}

\DeclareRobustCommand{\VAN}[3]{#2}
\let\VANthebibliography\thebibliography
\def\thebibliography{\DeclareRobustCommand{\VAN}[3]{##3}\VANthebibliography}

\usepackage{graphicx}	
\usepackage{amsmath}	
\usepackage[nameinlink, noabbrev]{cleveref} 

\crefname{equation}{Eq.}{Eqs.} 
\Crefname{equation}{Equation}{Equations} 

\crefname{figure}{Fig.}{Figs.}
\Crefname{figure}{Figure}{Figures}

\crefname{table}{Tab.}{Tabs.}
\Crefname{table}{Table}{Tables}

\crefname{section}{Sec.}{Secs.}
\Crefname{section}{Section}{Sections}
\crefname{appsec}{App.}{Apps.}
\Crefname{appsec}{App.}{Apps.}

\AddToHook{cmd/appendix/before}{%
  \crefalias{section}{appsec}%
}

\usepackage{enumitem}
\setlist[itemize]{leftmargin=*}

\title{Probing Quantum Gravity with Stage-IV Galaxy Surveys}

\title[GFT dark energy constraints]{
Constraining quantum-gravity predictions for evolving dark energy
}

\author[M. Tsedrik et al.]{
M. Tsedrik$^{1, 2}$\thanks{mtsedrik@ed.ac.uk}, 
B. Bose$^{1,2}$, 
L. Marchetti$^{3,4}$,
E. Ferreira$^{3}$ 
\\
$^1$Institute for Astronomy, University of Edinburgh,  Royal Observatory, Blackford Hill, Edinburgh, EH9 3HJ, UK\\
$^2$Higgs Centre for Theoretical Physics, School of Physics and Astronomy, Edinburgh, EH9 3FD, UK\\
$^3$Kavli Institute for the Physics and Mathematics of the Universe (WPI), UTIAS, The University of Tokyo, Chiba 277-8583, Japan\\
$^4$Okinawa Institute of Science and Technology Graduate University, Onna, Okinawa 904 0495 Japan
}

\date{Accepted XXX. Received YYY; in original form ZZZ}

\pubyear{\the\year{}}

\begin{document}
\label{firstpage}
\pagerange{\pageref{firstpage}--\pageref{lastpage}}
\maketitle

\begin{abstract}
 We confront a class of dark-energy equations of state emerging from group field theory (GFT) quantum gravity with DESI Data Release 2 baryon acoustic oscillations and Pantheon+ type-Ia supernovae. We introduce sampling parametrisations that replace microscopic initial-condition parameters by combinations more directly measured by background probes. The GFT solutions separate into logarithmic, power-law and oscillatory branches, determined by the microscopic interaction parameter $m$. The logarithmic branch is constrained to lie extremely close to a cosmological constant, while the power-law branch permits a small phantom deviation. Without perturbative-theory priors, oscillatory solutions can reproduce the mild preference of the distance data for a dip in $w(z)$ near $z\simeq0.5$--$1$. Profile-likelihood constraints favour $m\sim-2$ and $m\sim-5$ from BAO and supernovae, shifting towards $m\sim-3.5$ when CMB information is included. Conservative perturbative priors strongly suppress these deviations from $\Lambda$CDM. The quantum-gravity scale $z_q$, related to the average number of quantum gravity atoms, remains unconstrained, although its role in the time evolution makes higher-redshift observations a promising route to probing it. We further find that strong projection effects highlight the importance of performing likelihood profiling alongside our marginal posterior constraints. Our results provide a first direct test of GFT-motivated dynamical dark energy and demonstrate the potential for cosmological observations to inform quantum-gravity model building.
\end{abstract}



\section{Introduction}
\label{sec:introduction}

The new generation of cosmological surveys such as DESI~\footnote{\url{https://www.desi.lbl.gov}}, Euclid~\footnote{\url{https://www.euclid-ec.org}}, and Rubin~\footnote{\url{https://rubinobservatory.org}} will deliver Large-Scale Structure (LSS) measurements of unprecedented precision, offering new insights into the nature of the dark sector -- dark energy and dark matter. The statistical power and redshift reach of these Stage-IV surveys will enable percent-level constraints on the expansion history and growth of structure, opening a new window onto the physics driving cosmic acceleration. 

A key target of these surveys is the equation of state of dark energy, commonly parametrised through phenomenological forms such as the Chevallier-Polarski-Linder (CPL) $w_0$--$w_a$ form~\citep{Chevallier:2000qy,Linder:2002et}. While such parametrisations are useful for deriving model-independent constraints, they do not directly probe the underlying physical origin of dark energy~\footnote{Nevertheless, recent work has explored how observational constraints in the CPL plane can be mapped onto classes of scalar-field theories, providing a bridge between phenomenological dark-energy constraints and physically motivated models~\citep{Wolf:2023uno,Cataneo:2025vae,Garcia-Garcia:2026nzy}.}. Physically motivated models are essential for two complementary reasons. First, they come with theoretically grounded priors that restrict the allowed parameter space and render model comparison more informative. Second, posterior constraints on their parameters can feed back into the underlying theory, guiding and refining further model building. This interplay is especially important in quantum gravity, where using observations to discriminate among candidate theories and guide their development is both a central challenge and a long-standing goal. Late-time acceleration provides a particularly promising arena for this programme, as it may encode collective, low-energy manifestations of an underlying quantum-gravitational dynamics.

Recent developments in Group Field Theory (GFT) have provided a concrete step towards the realisation of this idea, with both inflation and late-time acceleration dynamically emerging from the collective behaviour of quantum geometric degrees of freedom~\citep{Marchetti:2025jze}. Group Field Theories (GFTs) are quantum and statistical field theories of fundamental spacetime quanta. They are part of the more broad framework of tensorial group field theories (TGFTs), which extend matrix and tensor models to a fully field-theoretic setting~\citep{Freidel:2005qe,Oriti:2006se,Oriti:2011jm,Carrozza:2013oiy,Carrozza:2016vsq,Gielen:2024sxs,Marchetti:2024tjq}. Their quantum-geometric degrees of freedom are encoded in group-theoretic variables, establishing close connections with canonical loop quantum gravity~\citep{Ashtekar:2004eh,Giesel:2011idc, Oriti:2013aqa,Oriti:2014yla}, spin foam models~\citep{Perez:2003vx,Perez:2012wv}, simplicial gravity path integrals~\citep{Bonzom:2009hw,Baratin:2010wi,Baratin:2011tx,Baratin:2011hp,Finocchiaro:2018hks}, and dynamical triangulations~\citep{Loll:1998aj,Ambjorn:2012jv,Jordan:2013sok,Loll:2019rdj}.

Effective, relational cosmological physics~\citep{Marchetti:2020umh,Marchetti:2024nnk} can be extracted from the GFT mean-field theory \citep{Marchetti:2022igl,Marchetti:2022nrf} (or, equivalently, in terms of appropriate condensate states), which provide a coarse-grained description of the underlying quantum gravitational dynamics~\citep{Gielen:2016dss,Pithis:2019tvp, Oriti:2016qtz,Gielen:2019kae,Marchetti:2020umh,Jercher:2021bie}. These models give rise to nonsingular bouncing cosmologies~\citep{Oriti:2016qtz,Gielen:2019kae,Marchetti:2020umh,Jercher:2021bie,Marchetti:2020qsq,Calcinari:2023sax}, as well as modifications of matter and perturbation dynamics~\citep{Ladstatter:2025kgu,Marchetti:2021gcv,Jercher:2023kfr,Jercher:2023nxa}. Of particular relevance here, GFT interactions can generate an effective late-time accelerating phase~\citep{deCesare:2016rsf,Oriti:2021rvm,Pang:2025jtk} with characteristic deviations from $\Lambda$CDM \citep{Marchetti:2025jze}, including oscillatory, logarithmic, or power-law modulations of $w(z)$ at late times, as well as the possibility of phantom behaviour, as preferred by the recent DESI results~\citep{DESI:2024mwx}. Ongoing extensions confirm the qualitative and partial quantitative robustness of the above results \citep{granata:acc}.

Given the precision of upcoming surveys, it is timely to ask whether such quantum gravity-motivated signatures can be tested observationally. In this work, we provide the first data constraints on a class of emergent dark energy models derived from interacting GFT dynamics. We focus on analytic forms of $w(z)$ obtained in \cite{Marchetti:2025jze}, providing constraints coming from a combination of background-only probes, specifically the Baryonic Acoustic Oscillations (BAO) feature and Supernovae type 1a (SN1a) measurements.

This paper is structured as follows. In \cref{sec:theory}, we summarise the cosmological and GFT theoretical framework, and the GFT effective parametrisations of $w(z)$. \cref{sec:data_priors} outlines the data sets and priors considered. In \cref{sec:results}, we present our results. We conclude in \cref{sec:conclusions}.


\section{Theory}
\label{sec:theory}

\subsection{Background observables}
\label{subsec:background_observables}

For a flat Universe with generic dark energy, the expansion rate is determined by
\begin{align}
        H(z)&=H_0E(z) \nonumber
        \\&=H_0\left[\Omega_{\rm bc}(1+z)^3+\Omega_\nu f_\nu(z)+\Omega_\gamma(1+z)^4+\Omega_{\rm de} f_{\rm de}(z) \right]^{1/2} \, , 
\end{align}
with the redshift-zero energy densities $\Omega_{i}$, where the index $i$ refers to baryons and cold dark matter (bc), neutrinos ($\nu$) photons ($\gamma$) and dark energy (de). The transition between relativistic and non-relativistic behaviour for massive neutrinos \citep{Lesgourgues:2006nd, Komatsu:2011} with the total mass $m_\nu$ is described by
\begin{equation}
    f_\nu = \frac{(1+z)^3}{1+\left(\frac{1+z}{6.328/m_\nu}\right)^{1/2}} \, , 
\end{equation}
fitted to CAMB \citep{camb}. The dark energy evolution is determined by
\begin{equation}
    f_{\rm de}(z) = \exp{\left[3\int_0^z \mathrm{d}z' \frac{(1+w(z'))}{1+z'} \right]}\, ,
\end{equation}
where $w(z)$ is the equation of state of dark energy.

From the expansion rate we can compute the transverse comoving distance
\begin{equation}
    D_M(z) = c \int_0^z \frac{d z'}{H(z')} ,
\end{equation}
while the radial Hubble distance is
\begin{equation}
    D_H(z) = \frac{c}{H(z)} .
\end{equation}
BAO measures these two scales normalised by the sound horizon at the baryon drag epoch, \(r_d\): $D_M(z)/r_d$ and $D_H(z)/r_d$. It is also sensitive to their geometric average $D_V(z)/r_d=(zD_M^2D_H)^{1/3}/r_d$. The sound horizon at the baryon drag epoch does not depend on dark energy, only on baryons, cold dark matter and neutrino content. 

For SNIa, the relevant observable is the distance modulus,
\begin{equation}
    \mu(z) = m_B(z) - M_B
    = 5 \log_{10} \left[
        \frac{D_L(z)}{10\,{\rm pc}}
    \right] \, , 
\end{equation}
where \(m_B\) and \(M_B\) are the apparent and absolute magnitudes. In a flat universe, the luminosity distance is given by 
\begin{equation}
    D_L(z) = (1+z)D_M(z) \, , 
\end{equation}
or, equivalently, for \(D_L\) expressed in Mpc,
\begin{equation}
    \mu(z)
    = 5 \log_{10} \left[
        \frac{D_L(z)}{{\rm Mpc}}
    \right] + 25 \, .
\end{equation}

Finally, in this work we will also be considering priors coming from the Cosmic Microwave Background (CMB) measurements of the Planck mission~\citep{planck2018cosmo}. The CMB measures the angular acoustic scale with great precision, corresponding to the sound horizon at recombination
\begin{equation}\label{eqn:thetastar}
    \theta_* = \frac{r_*}{D_M(z_*)}\, ,
\end{equation}
where the recombination redshift $z_* \approx 1090$ and the physical sound horizon
\begin{equation}
    r_*=\int_{z_*}^{\infty}\mathrm{d}z'\frac{c_s}{H(z')}\, .
\end{equation}
In this work, instead of using approximations~\citep[see e.g.][]{BOSS:2014hhw, Brieden:2022heh,Eisenstein:1997ik} we calculate $r_d$, $r_*$, and $z_*$ with CAMB for a $\Lambda$CDM cosmology. This follows from our assumption that the effective dark energy models considered in this paper do not affect early-time physics (see \cref{sec:data} for details). These properties are then used to derive the dimensionless Hubble constant $h=H_0/(100~\mathrm{km\, s}^{-1}\mathrm{Mpc}^{-1})$ from $\theta_*$ via a root-finding algorithm
\begin{equation}
    H_0 = \frac{1}{D_M(z_*)} c\int_0^{z_*}\frac{\mathrm{d}z'}{E(z')}=\frac{\theta_*}{r_*} c\int_0^{z_*}\frac{\mathrm{d}z'}{E(z')} \, . 
\end{equation}
While $r_*$ is not sensitive to late-time physics, $D_M(z_*)$ contains information on dark energy from low redshifts. Note that for this calculation we must assume a form for $w(z)$ across all redshifts up to the CMB. In the quantum gravity theories considered in this work, predictions for $w(z)$ are only robustly defined at low redshifts, but  corrections from $\Lambda$CDM are expected to be small at higher redshift (\cref{sec:data}). We therefore choose to assume a smooth transition to $w=-1$ at redshifts beyond our observations, $z>3$. Specifically, we apply a cosine taper to $1+w(z)$, which leaves the original equation of state unchanged below the transition redshift $z_{\rm t}$ and smoothly suppresses its deviation from $-1$ over a finite interval $\Delta z$, such that $w(z)=-1$ for $z\geq z_{\rm t}+\Delta z$. In our baseline analysis we adopt $z_{\rm t}=3$ and $\Delta z=\max(0.25,0.2z_{\rm t})=0.6$. We further study different transition redshifts and widths, as well as considering no transition at all~\footnote{We find that the choice of transition has negligible impact once the $\theta_*$ prior is included, except for the logarithmic model (see \cref{sec:models}), whose unsmoothed high-redshift extrapolation leads to a rapidly growing dark-energy density.}.

\subsection{Emergent dark energy from quantum gravity}\label{sec:models}

Within the GFT framework, cosmological dynamics emerges from the collective behaviour of quantum geometric degrees of freedom. At the effective level, the evolution of the Universe can be described by modified Friedmann dynamics sourced by an emergent fluid with equation of state $w(z)$, derived from the underlying mean-field quantum-gravitational dynamics \citep{Oriti:2016qtz,Marchetti:2020umh,Pang:2025jtk,Marchetti:2025jze}.

For a class of GFT models, the late-time cosmological evolution can be described in terms of an emergent dynamical dark energy component that generically approaches a de Sitter attractor characterised by $w \to -1$~\citep{Marchetti:2025jze}. Its phenomenological properties are encoded in the deviations $\delta w(z)$ from this fixed point:
\begin{equation}
w(z) = -1 + \delta w(z)\,.
\end{equation}
They depend on the interaction structure of the theory, and, within appropriate regimes, can be computed analytically. In such cases, $\delta w(z)$ takes the form
\begin{equation}
    \delta w(z)=\mathcal{T}(z;z_q)\mathcal{F}_m(z;z_q,A,B)\,,
\end{equation}
where $\mathcal{T}(z;z_q)$ is a universal suppression factor,
\begin{equation}
    \mathcal{T}(z;z_q)=[(1+z_q)/(1+z)]^{-6}\,,
\end{equation}
while $\mathcal{F}_m(z;z_q,A,B)$ is a model-dependent modulation. The above quantities 
depend on a small number of parameters \citep{Marchetti:2025jze}:
\begin{itemize}
    \item $m$: controls the phase structure of the GFT interactions and determines the functional behaviour of the modulation $\mathcal{F}_m$, see \cref{app:parametr}. We distinguish three classes of such modulations:
    \begin{enumerate}
        \item {Oscillatory} (
        $m< -4/3$);\label{item:osc}
        \item {Power-law} (
        $-4/3<m\le 0$); \label{item:pl}
        \item {Logarithmic} (
        $m=-4/3$).\label{item:log}
    \end{enumerate}
    The value $m=-6$ is of particular theoretical interest, as it corresponds to interactions which can be motivated by discrete-gravity path integrals \citep{Ladstatter:2025kgu,Marchetti:2025jze}. We will often refer to models with $m=-6$ as \qmarks{quantum geometric} in the following.
    \item $z_q$: a characteristic redshift related to the average number  $\mathcal{N}(z)$ of quantum gravity atoms 
    \begin{equation}\label{eqn:nofz}
        \mathcal{N}(z)=\left[\frac{1+z_q}{1+z}\right]^3\,.
    \end{equation}
    In particular, today $\mathcal{N}_0\equiv\mathcal{N}(z=0)=(1+z_q)^3$, making $z_q$ a parameter with a genuine quantum-gravitational interpretation.
    \item $A$, $B$: amplitudes set by initial conditions of the mean-field GFT dynamics.
\end{itemize}
Among these parameters, $m$ provides the most direct window into the microscopic physics by characterizing the underlying quantum-gravitational interactions. The parameter $z_q$, meanwhile, probes the quantum-geometric content of the cosmological state through the genuinely quantum observable $\mathcal{N}_0$. Both quantities differ conceptually from $A$ and $B$, which encode initial conditions and therefore carry less direct microscopic information. For these reasons, $m$ and $z_q$ will be the primary quantities of physical interest in our analysis below.

Although $\{m,z_q, A,B\}$ provide a phenomenologically natural parametrisation of the emergent dark energy equation of state, below we adopt a new parametrisation of $\delta w(z)$ which is better constrained by the background probes utilised in this work (see \cref{sec:data}). 

\begin{figure*}
    \centering
    \includegraphics[width=0.95\textwidth]{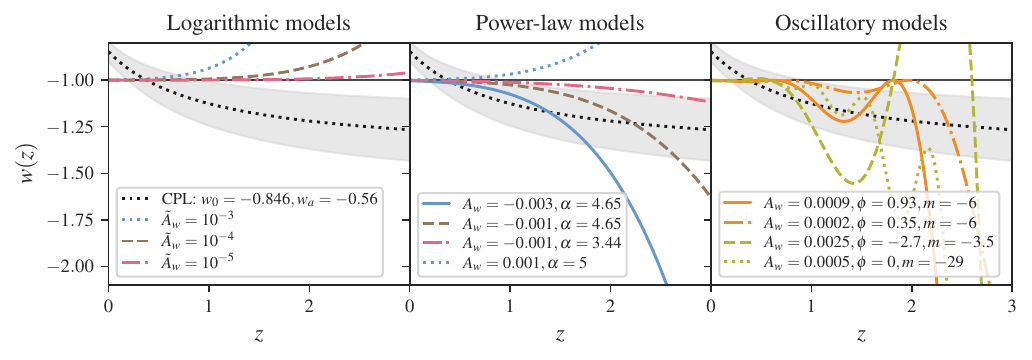}
    \includegraphics[width=0.95\textwidth]{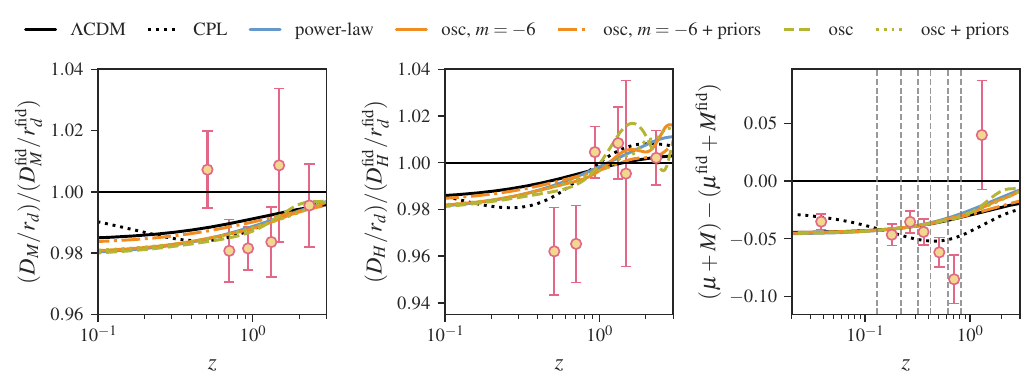}
\caption{{\bf Top panels:} Representative $w(z)$ evolution for the oscillatory, logarithmic, and power-law quantum gravity–inspired models. Coloured curves illustrate characteristic behaviour within each family, while the dotted curve shows the DESI best-fit CPL model. The grey band indicates the corresponding $1\sigma$ uncertainty from the DESI errors on $(w_0,w_a)$. {\bf Bottom panels:} The DESI DR2 BAO (perpendicular and parallel) and Pantheon+ SN (distance modulus) measurements and theoretical predictions at the best-fit values from \cref{tab:constraints} normalised to the CMB $\Lambda$CDM model predictions. The dashed vertical lines in the SN plot denote the re-binning of the SNe in redshift for visual purposes only.}
    \label{fig:wz_models}
\end{figure*}

\subsubsection{Sampling parametrisations} \label{sec:samp_param}

Background data more directly constrain an amplitude $A_w$ and time-evolution $f_w(z)$ of $\delta w(z)$,
\begin{equation}\label{eqn:newpara}
    \delta w(z)=A_wf_w(z)\,, 
\end{equation}
where $A_w \propto (1+z_q)^{-6}$. Further, for a given dataset, the best measured parameter is the equation of state at some pivotal redshift $w_{\rm piv} \equiv w(z_{\rm piv})$~\citep{Cortes:2024lgw}. Hence, the amplitude $A_w$ can be found by 
\begin{equation}
A_w=(w_{\rm piv}+1)/f_w(z_{\rm piv}) \, , 
\end{equation}
for all parameter combinations with $f_w(z_{\rm piv}) \neq 0$. Due to these singular points in parameter space for the amplitude, we choose to sample directly in $A_{w}$ rather than $w_{\rm piv}$. 

Below we provide a parametrisation of the models \ref{item:osc}-\ref{item:log} based on \cref{eqn:newpara}.
\paragraph*{\ref{item:osc} Oscillatory models:} These can be parametrised entirely in terms of three parameters: $\{A_w, \beta,\phi\}$, with 
\begin{equation}\beta=\beta(m)=2\sqrt{-1-3m/4}\,,
\end{equation}
$\phi=\phi(m,z_q,A,B)$, and $A_w=A_w(m,z_q,A,B)$. The time-dependent behaviour of $\delta w(z)$ is captured by
\begin{equation}\label{eqn:oscnewpara}
    f_w(z;\beta,\phi)=(1+z)^6\left[c(\beta)+\cos(3\beta\log(1+z)+\phi)\right]\,.
\end{equation}
We refer the reader to \cref{app:parametr} for details. In the limit \(\beta \rightarrow 0\) (\(m\rightarrow -4/3\)), the logarithmic models are recovered, as expected. Moreover, standard cosmology is recovered for \(A_w=0\), corresponding to \(A,B\rightarrow 0\), or equivalently \(z_q\rightarrow \infty\), see \cref{eqn:parametersoscnew}.

By modelling this form in polar coordinates, the $\Lambda$CDM limit is fully specified by $A_w=0$, independently of $\phi$. Thus, at the $\Lambda$CDM limit the phase is unidentifiable, leaving an unconstrained direction in the polar parametrisation. Moreover, because $\phi$ is periodic, sampling requires choosing a fundamental interval, e.g.\ $\phi\in[0,2\pi)$.

We therefore re-chart the model in Cartesian coordinates,
\begin{equation}
X=A_w\cos\phi,
\qquad
Y=A_w\sin\phi,
\end{equation}
for which the $\Lambda$CDM limit corresponds uniquely to $(X,Y)=(0,0)$. We retain the polar variables for interpreting the oscillatory equation of state and, when the amplitude is sufficiently large that the phase is identifiable, for presenting profile-likelihood results, since these are independent of the prior measure.

Importantly, however, a uniform prior in $(X,Y)$ does not induce a uniform prior in $(A_w,\phi)$. Since
\begin{equation}
\mathrm{d}X,\mathrm{d}Y
\propto
A_w\mathrm{d}A_w \, , 
\end{equation}
the induced prior on the amplitude introduces an informative prior on the posterior as
\begin{equation}
\label{eq:a_w_from_x_y}
P(A_w)\mathrm{d}A_w
\propto
A_w\mathrm{d}A_w \, .
\end{equation}
Consequently, the marginal posterior density in $A_w$ necessarily vanishes as $A_w\rightarrow0$, even if the likelihood itself is maximised at the $\Lambda$CDM limit. Marginal posterior constraints on $A_w$ can therefore exhibit an apparent preference away from zero purely as a prior-volume (projection) effect.

\paragraph*{\ref{item:pl} Power-law models:} These models can be parametrised in terms of $\{A_w,\alpha\}$, where $A_w\equiv A_w(m,z_q,B)$, and 
\begin{equation}
    {\alpha}={\alpha}(m)=6(1-\sqrt{1+3m/4}) \, .
    \label{eq:alpha_of_m}
\end{equation}
The time-dependent behaviour of $\delta w(z)$ is captured by
\begin{equation}\label{eqn:timedeppl}
    f_w(z;\alpha)=(1+z)^{{\alpha}}\,.
\end{equation}
The sign of $A_w$ (but not its amplitude) is fixed by the value of $\alpha$, $\mathrm{sgn}(A_w)=\mathrm{sgn}(\alpha-3(4-\sqrt{6}))$. Standard cosmology is recovered in the limit $A_w\to 0$, which occurs for \(z_q\rightarrow\infty\), for \({\alpha}=3(4-\sqrt{6})\), and for $\alpha=0$.

\paragraph*{\ref{item:log} Logarithmic models:}
These models can be parametrised in terms of $\{A_w,z_q\}$, with $A_w=A_w(B,z_q)\ge 0$. The time-dependent behaviour of $\delta w(z)$ is given by
\begin{equation}\label{eqn:timedeplog}
    f_w(z;z_q)=(1+z)^6\log^2\left(\frac{1+z_q}{1+z}\right) \, .
\end{equation}
Thus, the equation of state transitions to \(w=-1\) at
$z = z_q$ . For \(z>z_q\), the logarithmic contribution grows, leading to \(w(z)>-1\). For the observational redshift range considered here, \(0<z<3\), theoretical priors (see \cref{eqn:structuralprior} below) imply \(z_q \gg z\). The logarithmic term therefore makes only a negligible contribution to the time evolution over the redshifts probed by the data. We approximate this limit by treating the logarithmic models as a sub-class of the power-law models characterised by an amplitude $\tilde{A}_w$ given by
\begin{equation}
    \widetilde{A}_w= A_w \log^2(1+z_q)
\geq 0 \,,
\end{equation}
where, on the right-hand side, $A_w$ refers to the amplitude in the corresponding logarithmic model, and taking the limit
\begin{equation}
{\alpha} \to 6 \, (m\to-4/3)\,.
\end{equation}
\vspace{-5mm}
\par\noindent\hfill$\diamond$\hfill\mbox{}\par
\vspace{2mm}
An important consequence of these parametrisations is that much of the dependence on $z_q$ that enters through the overall amplitude in the original parameters is absorbed into the independent parameter $A_w$. Constraints on $z_q$ therefore arise primarily through its residual effect on the redshift dependence $f_w(z)$, which can be weak over the observational range and consequently makes $z_q$ difficult to constrain. For the oscillatory model, the additional dependence of $z_q$ through the oscillation phase is similarly absorbed into the independent parameter $\phi$, further reducing the sensitivity to $z_q$ (see \cref{app:parametr}). 

In the top panels of \cref{fig:wz_models}, we show representative evolutions of the equation of state for the logarithmic, power-law and oscillatory models, together with $\Lambda$CDM and the DESI best-fit CPL model~\citep{DESI:2025zgx}. The logarithmic model permits only $w(z)>-1$, and is therefore unable to reproduce the phantom evolution favoured by the DESI CPL fit. The power-law models allow both signs of the deviation from $\Lambda$CDM and can closely reproduce a CPL-like evolution over the redshift range probed by current observations. The oscillatory model exhibits considerably greater freedom, with its amplitude, phase and frequency allowing both crossings of the phantom divide and non-monotonic evolution that cannot be captured by the CPL parametrisation.

The best-fit model predictions for the DESI DR2 BAO and Pantheon+ SN observables are shown in the bottom panels of \cref{fig:wz_models}, normalised to the CMB $\Lambda$CDM prediction (see \cref{tab:constraints}). Despite the substantially different instantaneous $w(z)$ evolutions permitted by these models, their effects on the distance observables are considerably smoother, reflecting the integrated dependence of cosmological distances on the expansion history. In particular, oscillatory features in $w(z)$ can be strongly suppressed in the BAO and SN observables, allowing qualitatively different dark-energy histories to produce similar predictions over the observed redshift range.

\begin{table*}
\centering
\caption{
Quantum-gravity model priors on the sampling parameters defined in \cref{sec:samp_param}. The theoretical consistency priors described in \cref{sec:theoryprior} are imposed in addition to the sampling ranges below.
}
\label{tab:qg_priors}
\begin{tabular}{llll}
\toprule
Model & Parameter & Prior & Notes \\[.5mm]
\hline
\midrule
Logarithmic
& \(\widetilde A_w\)
& \(\mathcal{U}[0., 0.01]\)
& \(\widetilde A_w \geq 0\) \\

& \(\log_{10} z_q\)
& \(\mathcal{U}[\log_{10}3, 5]\)
& Sampled when imposing the technical \(A,B\) prior \\

\midrule

Power law
& \(A_w\)
& \(\mathcal{U}[-0.06, 0.06]\)
& Amplitude \\

& \(\alpha\)
& \(\mathcal{U}[0., 6)\)
& Power-law index \\

& \(\log_{10} z_q\)
&  \(\mathcal{U}[\log_{10}3, 5]\)
& Sampled when imposing the technical \(A,B\) prior \\

\midrule

Oscillatory

& \(\beta\)
& \(\mathcal{U}[0, 12]\)
& Oscillation parameter: frequency \\

& \(X\)
& \(\mathcal{U}[-0.05, 0.05]\)
& Amplitude and phase: $X= A_w \cos{\phi}$ \\

& \(Y\)
& \(\mathcal{U}[-0.05, 0.05]\)
& Amplitude and phase: $Y= A_w \sin{\phi}$ \\


& \(\log_{10} z_q\)
&  \(\mathcal{U}[\log_{10}3, 5]\)
& Sampled when imposing the technical \(A,B\) prior \\

\midrule

Theory prior
& \(z_{\rm in}\)
& \(3\)
& Maximum redshift over which the late-time description is required \\

& \(M_{\rm safe}\)
& \(0.1\)
& Fiducial safety factor for the technical \(A,B\) prior \\

\bottomrule
\end{tabular}
\end{table*}

\begin{table}
\centering
\caption{Cosmological priors adopted in our analyses.} 
\label{tab:cosmo_priors}
\begin{tabularx}{\linewidth}{lXX}
\toprule
Parameter & Prior & Notes \\[.5mm]
\hline
\midrule
\(\Omega_{\rm m}\) &
\(\mathcal{U}[0.05, 0.5]\) &
Total matter density \\

\(h\) &
\(\mathcal{U}[0.5, 0.9]\) &
Expansion rate \\

\(\omega_{\rm b}\) &
\(\mathcal{U}[0.005, 0.1]\) &
Baryon density \\
 &
\(\mathcal{N}(0.02223, 0.000146)\) &
{\it CMB prior} \\

\(\omega_{\rm bc}\) &
 
 \(\mathcal{N}(0.14208,  0.00122)\) &
Baryon + CDM density \\
 &
&
{\it  CMB prior} \\

\(\theta_\star\) &
\(\mathcal{N}(1.04110, 0.000257)\)
&
Acoustic scale \\
 &
&
{\it  CMB prior} \\

\(\sum m_\nu\) &
\(0.06\,{\rm eV}\) &
Fixed \\

\midrule
\(w_0\) &
\(\mathcal{U}[-2,-0.3]\) &
CPL only \\

\(w_a\) &
\(\mathcal{U}[-2, 2]\) &
CPL only \\
\bottomrule
\end{tabularx}
\end{table}

\subsection{Theoretical Priors}
\label{sec:theoryprior}

Theoretical priors are essential when confronting fundamental models with observations, since theoretical consistency can restrict both the allowed values and combinations of their parameters. Ignoring these restrictions enlarges the parameter space to include configurations that cannot be realised by the underlying theory, potentially altering parameter constraints, best-fit predictions, Bayesian model comparison and potentially false detections of new physics. Deriving such restrictions from the underlying theory therefore is not only necessary for robustly testing the theory, but further provides information that is absent from purely phenomenological parametrisations.

The extraction of macroscopic, homogeneous cosmological dynamics from the underlying GFT model of quantum gravity relies on several approximations that restrict the parameter space of the models introduced in \cref{sec:models}. The most important is the mean-field approximation, which is expected to be reliable in the large-field, or equivalently large-occupation-number, regime \citep{Marchetti:2020qsq}. Relative quantum-gravity fluctuations are then suppressed, allowing a semiclassical description to emerge\footnote{This regime is also expected to support the single-mode approximation adopted here \citep{Marchetti:2020qsq,Gielen:2016uft}.} \citep{Marchetti:2020qsq,Gielen:2019kae}. Systematically going beyond this regime requires a level of control over the full quantum gravity theory that is currently beyond reach. We therefore regard the requirement of a large GFT mean field as a structural consistency condition and impose the corresponding theoretical priors.

By contrast, some of the conditions entering the derivation of the analytical expressions in \cref{sec:models} arise from a perturbative approximation rather than from physical consistency of the underlying model. As shown in \cite{Marchetti:2025jze}, non-perturbative numerical solutions closely reproduce the perturbative dynamics beyond these conditions. We therefore distinguish these approximation-validity requirements from genuine theoretical priors and interpret constraints arising from them with appropriate caution.

The theoretical conditions are most naturally formulated in terms of the
underlying GFT parameters $(A,B,m,z_q)$, whereas our inference is performed
using the sampling parametrisation introduced in
\cref{sec:samp_param}. We therefore impose the priors below on
the corresponding underlying parameters at each sampled point in our analyses. Their
mapping into the sampling variables is in general non-trivial: in
particular, since $A_w$ depends jointly on $A$, $B$ and $z_q$, the resulting
theoretical prior on $A_w$ is correlated with the parameters controlling
the time dependence in oscillatory models. 

Below we discuss in detail theoretical priors induced by the above structural and technical constraints.

\subsubsection{Structural priors: $z_q$}
The redshift $z_q$ depends on the elementary quantum volume $\mathfrak{v}$ and the fiducial volume $V_0$~\citep{Marchetti:2020umh} via~\citep{Marchetti:2025jze}
\begin{equation}
1+z_q=[V_0/\mathfrak{v}]^{1/3}\,.
\end{equation}
However, since the value of $\mathfrak{v}$ is not predicted by the theory, it is difficult to motivate a specific theoretical prior on $z_q$. Even assuming $\mathfrak{v}$ to be of Planckian size would require specifying a prior for the fiducial volume $V_0$; identifying $V_0$ with the observable volume today would be circular, since the latter is dynamically generated by the GFT evolution, which itself depends on $z_q$. The only robust theoretical requirement is therefore that the average number of quantum-gravity atoms, $\mathcal{N}(z)$, be large at the earliest observational redshift $z_{\mathrm{in}}$, namely
\begin{equation}\label{eqn:structuralprior}
\mathcal{N}_{\mathrm{in}}
\equiv\mathcal{N}(z_{\mathrm{in}})
=\left(\frac{1+z_q}{1+z_{\mathrm{in}}}\right)^3
\gg 1\quad\longrightarrow\quad z_q\gg z_{\mathrm{in}}\,.
\end{equation}
For the observations used in this work, we take $z_{\rm in}=3$. Since the theory does not provide a quantitative threshold for the condition $\mathcal{N}_{\rm in}\gg1$, we impose only the minimal bound $z_q>z_{\rm in}=3$.

\subsubsection{Technical priors: $A$ and $B$}
For the initial data $A$ and $B$, the relevant conditions come from the requirement that the perturbative regime used to derive the analytic expressions for $\delta w(z)$ is satisfied. The validity of the perturbative regime can be expressed as $\vert\xi(z)\vert\ll 1$, where $\xi(z)$ satisfies \citep{Marchetti:2025jze}
\begin{equation}\label{eqn:deltax}
   \mathcal{N}(z) \xi(z)=\begin{cases}
        A\mathcal{N}^{-\mu}(z)+B\mathcal{N}^{\mu}(z)\,,\quad &0<\mu^2\le 1\,,\\
        A+(B/2)\log\mathcal{N}(z)\,,\quad &\mu^2=0\,,\\
        A\cos\Phi(z)+B\sin\Phi(z)\,,\quad &\mu^2<0\,,
    \end{cases}
\end{equation}
 where $\mu=\sqrt{1+3m/4}$ (hence $\mu^2=-\beta^2/4$) and $\Phi(z) = (3\beta/2) \log[(1+z_q)/(1+z)]$.
Clearly, $\mathcal{N}(z)\in[\mathcal{N}_{\min},\mathcal{N}_{\max}]$, with $\mathcal{N}_{\max}=(1+z_q)^3$, and $\mathcal{N}_{\min}= \mathcal{N}_{\text{in}}=[(1+z_q)/(1+z_{\text{in}})]^3$ being the average number of quantum gravity atoms today and at the earliest observational redshifts, respectively. The structural requirement motivates $\mathcal{N}_{\rm in}\gg1$~(\cref{eqn:structuralprior}), although, as discussed above, in practice we impose only the minimal condition $\mathcal{N}_{\rm in}>1$. The perturbative condition $\vert\xi(z)\vert\ll 1$ then requires 
\begin{subequations}
\label{eq:priors}
\begin{equation}
    \vert B\vert\ll\begin{cases}
        \mathcal{N}_{\text{in}}^{1-\mu}\,, &0<\mu^2\le 1\, \quad ({\rm PL})  \\
        2\mathcal{N}_{\text{in}}/\log\mathcal{N}_{\text{in}}, &\mu^2=0\, \quad ({\rm Log})
    \end{cases}\,, \, 
\end{equation}
the values of $A$ being irrelevant, as it controls the subdominant modes for $w(z)$ for $0\le \mu^2\le 1$. On the other hand, for $\mu^2<0$,
\begin{equation}
    \vert A\vert\ll \mathcal{N}_{\text{in}}\,,\quad \vert B\vert\ll \mathcal{N}_{\text{in}}\,,\,\quad \mu^2<0 \quad \text{(Osc)}\,.
\end{equation}
\end{subequations}
In practice we impose the following prior
\begin{equation}
   \vert A\vert\leq M_{\rm safe} A_{\rm max} \, , \quad \vert B\vert\leq M_{\rm safe} B_{\rm max} \, , 
   \label{eq:msafe_prior}
\end{equation}
where $A_{\rm max}$ and $B_{\rm max}$ are taken to be the limits above, dependent on $z_q$. $M_{\rm safe}$ is a constant to be chosen, which we set to $0.1$ (see \cref{sec:data_priors}). 

We remark again that the above conditions ensure the validity of the perturbative regime underlying the explicit equations of ~\cref{sec:models}. Nevertheless, the close agreement between the perturbative and non-perturbative quantum-gravity dynamics suggests that these equations may remain reliable beyond this regime, motivating us also to consider the case in which the corresponding priors are not imposed.


\section{Data and Priors}
\label{sec:data_priors}

\subsection{Data}\label{sec:data}
When confronting the models introduced in \cref{sec:models} with observations, two limitations must be considered. First, the expressions for $w(z)$ presented there apply only in the late-time, large-volume regime, and extending the description to earlier times would require solving the full mean-field dynamics, which is computationally challenging. Nevertheless, their qualitative structure suggests that departures from $\Lambda$CDM should become progressively less relevant at high redshift. Indeed, the GFT interactions responsible for the emergent dark-energy dynamics contain an overall factor responsible for the cosmological-constant-like behaviour, modulated by bounded oscillatory terms \citep{Marchetti:2025jze}. Since the contribution of the entire interaction sector to the total energy density becomes increasingly subdominant at high redshift, these modulations are expected to induce only small corrections to the background expansion. This motivates adopting $\Lambda$CDM as a minimal high-redshift completion of the model.

The second limitation concerns cosmological perturbations. The models presently describe only the homogeneous background, while a comprehensive framework for extracting the dynamics of inhomogeneities from GFT is still lacking, an intrinsically difficult problem in any fully non-perturbative approach to quantum gravity, although pioneering results exist for restricted classes of observables and non-interacting regimes \citep{Jercher:2023kfr,Jercher:2023nxa}. 

Because of these limitations, we restrict our baseline analysis to late-time probes whose theoretical predictions depend only on the background dynamics.
More precisely, the datasets we consider are:
\begin{itemize}
    \item Supernovae type 1a (SN1a) measurements from the Pantheon Plus data set~\citep{Scolnic:2021amr,Brout:2022vxf}.
    \item The DESI DR2 Baryon Acoustic Oscillations (BAO) measurements~\citep{DESI:2025zgx}.
\end{itemize}
\subsection{Priors}
We adopt broad uniform priors on the late-time cosmological and
dark-energy parameters, summarised in \cref{tab:cosmo_priors}.
All analyses assume spatial flatness and fix the sum of neutrino masses
to \(\sum m_\nu = 0.06\,{\rm eV}\). For BAO+SNIa-only runs we sample
\((\Omega_{\rm m},h)\) directly. When including early-Universe
information, we impose Gaussian CMB priors that are insensitive to the
late-time expansion history on the physical baryon and baryon+CDM
densities, \(\omega_{\rm b}\equiv\Omega_{\rm b}h^2\) and
\(\omega_{\rm bc}\equiv\Omega_{\rm bc}h^2\), following
\citet{Lemos:2023xhs}. 

In a separate analysis, we additionally impose a CMB prior on the angular acoustic scale, $\theta_*$ (\cref{eqn:thetastar}). Its theoretical value is entirely determined by the background expansion history, together with standard pre-recombination physics, and can therefore be computed without specifying any perturbation dynamics. Evaluating it nevertheless requires specifying the background beyond $z\sim3$, up to which we regard the parametrisation of \cref{sec:models} as reliably predictive. For the reasons discussed in \cref{sec:data}, we adopt a minimal completion of the models of \cref{sec:models} in which the background is approximated by $\Lambda$CDM beyond this range and standard physics is recovered around recombination.\footnote{We assume that quantum-gravity effects do not appreciably modify the microphysics of recombination or the perturbation dynamics governing the acoustic scale. Generic quantum-gravity corrections are expected to be negligible at the corresponding energy scales, while the GFT interaction terms responsible for the modified dark-energy dynamics are themselves strongly suppressed at these redshifts, as discussed in \cref{sec:data}.}

Priors for the parameters of the models in \cref{sec:models} are given in \cref{tab:qg_priors}. In addition to these sampling priors, we impose the theoretical priors discussed in \cref{sec:theoryprior}. We take \(z_{\rm in}=3\), a conservative choice encompassing the highest-redshift DESI BAO measurement at \(z=2.33\). We consider analyses both with and without the technical \(A\) and \(B\) priors. Unless otherwise stated, we adopt a fiducial value of \(M_{\rm safe}=0.1\) in \cref{eq:msafe_prior}. We emphasise that these are theoretical consistency conditions rather than observational priors.

\section{Results} 
\label{sec:results}

For our analyses we make use of the \href{https://github.com/joezuntz/cosmosis-standard-library}{\texttt{CosmoSIS}} package, in which we include a new module which calculates the background quantities under the different quantum gravity models. This module will be made publicly available upon publication. We use \texttt{Nautilus}~\citep{Lange:2023ydq} with 5,000 live points and have cross-checked this for our baseline runs with the \texttt{PolyChord}~\citep{Handley:2015fda} sampler with excellent agreement. To complement the marginal posterior constraints and assess the impact of marginalisation of the highly non-Gaussian posteriors, we additionally profile likelihoods and posteriors using \href{https://github.com/scikit-hep/iminuit}{\texttt{iminuit}}~\citep{iminuit}, with confidence intervals determined using the MINOS algorithm. With such profiles we obtain a one-dimensional distribution by evaluating the posterior over
a scan of a parameter of interest while maximising the
posterior with respect to all other parameters. A difference in 1D constraints between integration and maximisation of the posterior is approximately given by the so-called Laplace term, which accounts for volume effects as explained in ~\citet{Hadzhiyska:2023wae, Tsedrik:2025hmj}. We additionally compute the best-fit values with least-squares optimisation using the Migrad algorithm. We use the best-fit metric, i.e. $\chi^2(\theta_{\rm MAP})$, computed at the maximum a posteriori (MAP) point with CMB priors. Due to the complex set of degeneracies in the parameter space and the sensitivity to prior choices, as well as sampling parametrisations, we prefer this metric for model comparison over Bayesian evidence.

These results will provide a first quantitative assessment of the potential of large-scale structure surveys to probe quantum gravity through the dynamics of dark energy. We also investigate the impact of the theoretical priors discussed in \cref{sec:theoryprior} on these results. All our results are presented in \cref{tab:constraints}.

\begin{figure}
    \centering
 \includegraphics[width=0.8\columnwidth]{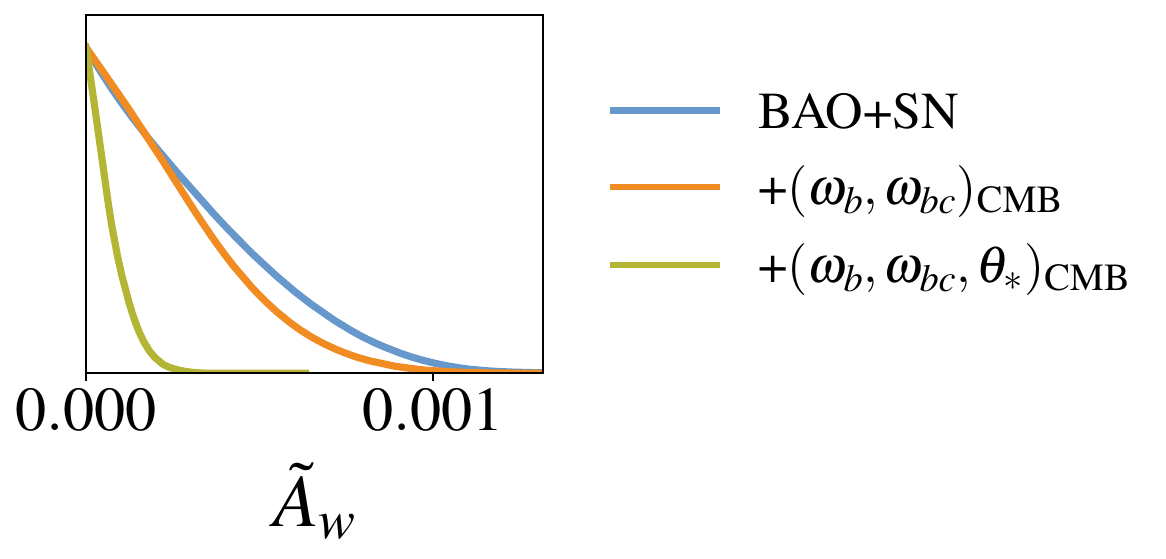}
  \includegraphics[width=0.9\columnwidth]{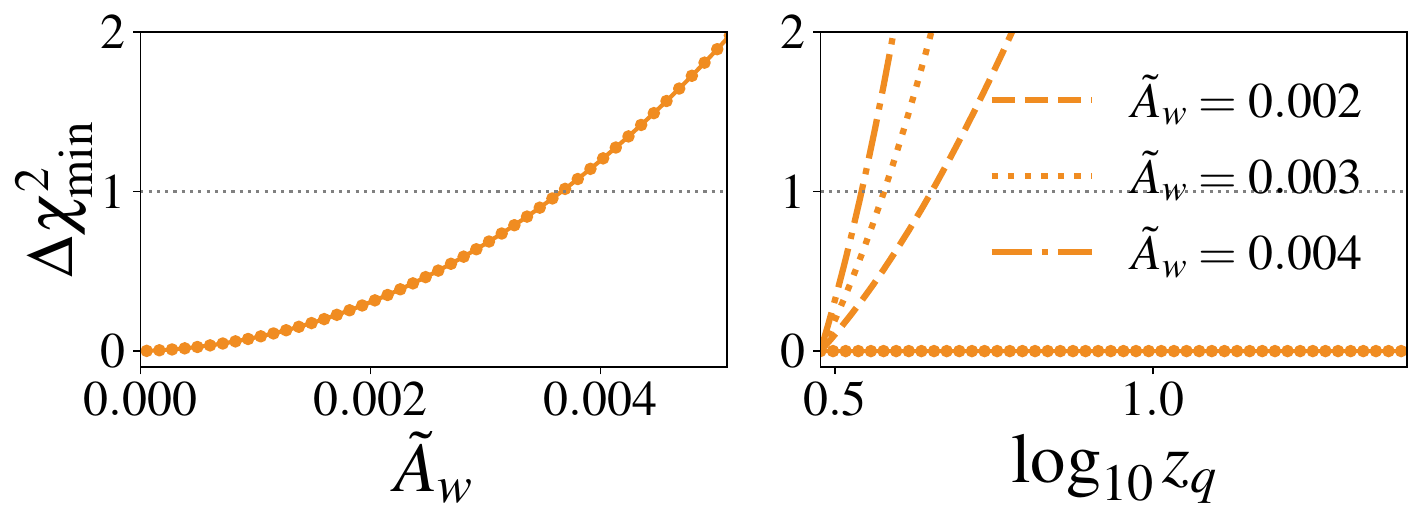}
    \caption{Logarithmic model results. {\bf Top panel:}  marginalised constraints on the dark-energy amplitude $\tilde{A}_w$ for the three combinations of data specified in the legend. {\bf Bottom panel:} posterior profiling  in $\tilde{A}_w$ and $z_q$ for BAO+SN+$(\omega_{\rm b}, \omega_{\rm bc})_{\rm CMB}$.}
    \label{fig:log_model_contours}
\end{figure}

{\bf\textit{Logarithmic models}} introduce a characteristic $w(z)>-1$ behaviour at $z>0$, which is disfavoured by the data. In the upper panel of \cref{fig:log_model_contours} we show constraints on the amplitude parameter $\tilde{A}_w$, which is consistent with its $\Lambda$CDM limit with and without CMB information. Cosmological parameters stay consistent in these data-combinations as well, e.g. $\Omega_{\rm m}=0.3003 \pm 0.0085$ for BAO+SN and  $0.3025\pm 0.0035$ with the inclusion of CMB information. We then re-introduce the $\log_{10}z_q$ dependence in $f_w(z) \propto \log^2((1+z_q)/(1+z))$ and vary this parameter under the consideration of the technical prior in \cref{eq:msafe_prior}. This introduces a hard bound on the maximum allowed amplitude:
\begin{equation}
    A_{w, \rm max}= \frac{16 M^2_{\rm safe}}{3 (1+z_{\rm in})^6}\, ,
\end{equation}
but does not change the conclusion on the preference for the $\Lambda$CDM limit. In the lower panel of \cref{fig:log_model_contours} we show the profiling of the posterior in one combination of datasets for this model, SN+BAO+CMB-densities without $\theta_*$. The inclusion of $\theta_*$ makes the result sensitive to smoothing, as dark energy densities increase rapidly at high redshifts for this form of $w(z)$. The amplitude parameter's $1\sigma$ region is increased with respect to the posterior in the upper panel because we have an additional degree of freedom; its best-fit is consistent with zero. This means that when profiling $\log_{10} z_q$ the amplitude is zero, which results in a flat profile in $z_q$, i.e. the parameter is unconstrained. 

This being said, if our data preferred a different Universe with a non-zero amplitude, we could detect dependence on the quantum-gravity parameter $z_q$ through the time-evolution in $f_w(z)$. We demonstrate this by profiling $\log_{10}z_q$ for fixed non-zero amplitude values. The preference for $\Lambda$CDM in the posterior distribution can be misleading upon marginalisation: by sampling in $(\tilde{A}_w, \log_{10}z_q)$, the marginalised 1D $z_q$ distribution will peak around its theoretical limit $z_q=z_{\rm in}$, while by sampling in $(B, \log_{10}z_q)$, due to the projected large volume in small $B$ and large $z_q$, one will obtain an apparent preference for $z_q>10$. Both of these conclusions are due to projection effects as we know the likelihood is flat in $z_q$ for this combination of data.

\begin{figure}
    \centering
 \includegraphics[width=0.7\columnwidth]{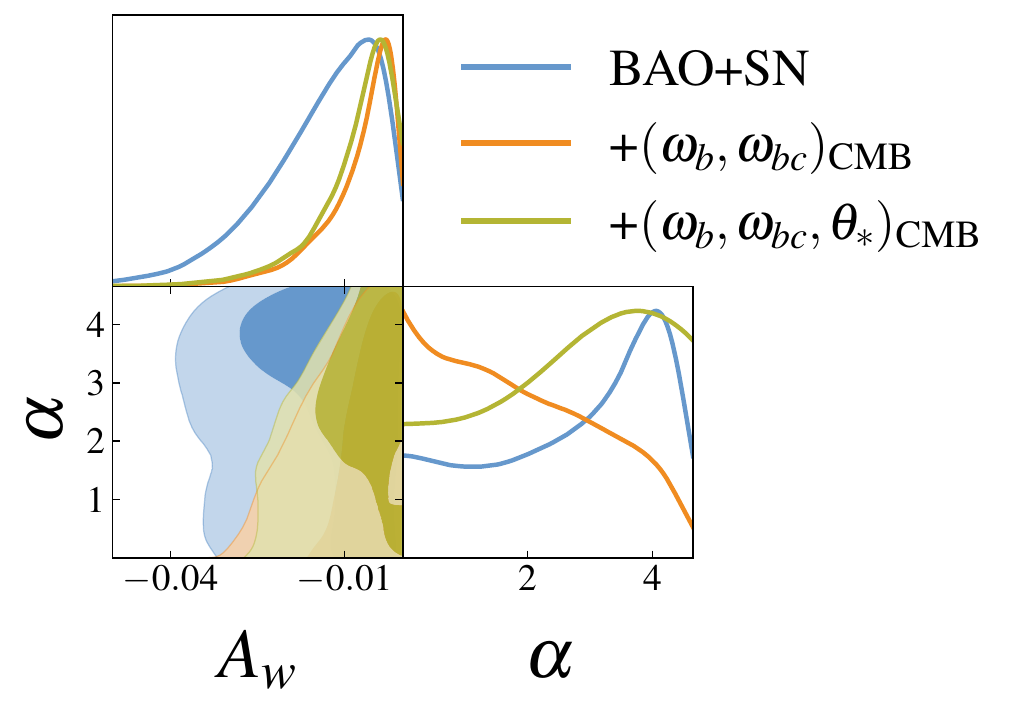}
  \includegraphics[width=0.9\columnwidth]{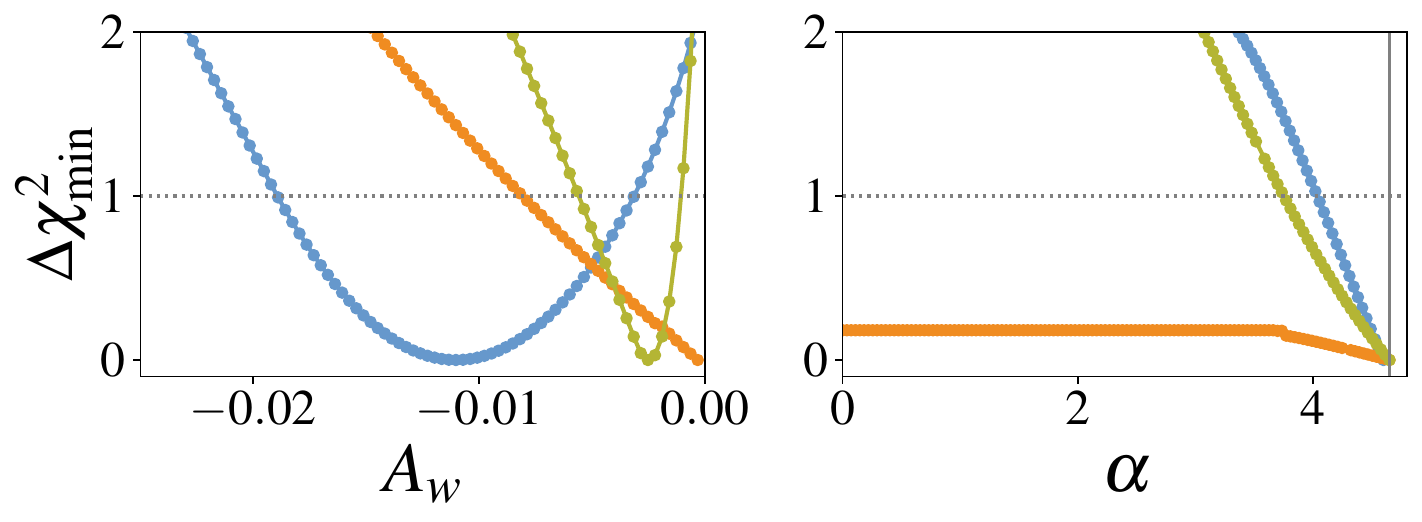}
  \caption{Power-law model results. {\bf Top panel:}  marginalised constraints on the dark-energy amplitude $A_w$ and evolution-controlling exponent $\alpha$ for the three combinations of data specified in the legend. {\bf Bottom panel:} posterior profiling  with the same colour-coding for the datasets. Both results are without technical priors.}
    \label{fig:pl_model_contours}
\end{figure}

\begin{figure}
    \centering
  \includegraphics[width=0.45\columnwidth]{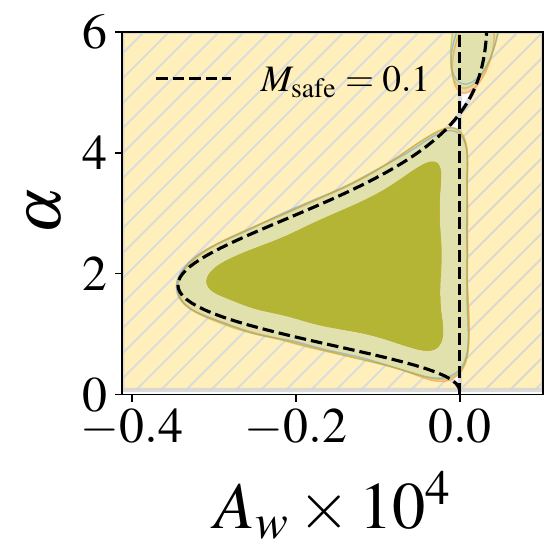}
   \includegraphics[width=0.45\columnwidth]{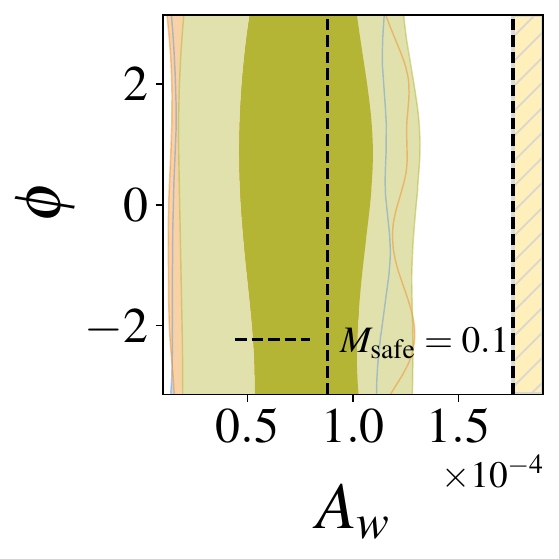}
\caption{The impact of technical priors on the constraints within  power-law ({\bf left panel}) and oscillatory $m=-6$ ({\bf right panel}) models.}
    \label{fig:pl_model_contours_priors}
\end{figure}

{\bf \textit{Power-law models}} can be separated into two branches around its $\Lambda$CDM limit \footnote{For a similar ``sign-switching'' behaviour see an interacting dark energy model with pure momentum exchange \citep{Pourtsidou:2016ico, Carrilho:2022mon, Tsedrik:2025jdv}.}: negative amplitude $A_w$ and $\alpha<3(4-\sqrt{6})$ with phantom behaviour, positive amplitude $A_w$ and $\alpha>3(4-\sqrt{6})$ with $w(z)>-1$ similar to the logarithmic models above. The datasets considered in this analysis prefer phantom behaviour, which is illustrated in \cref{fig:pl_model_contours} via a clear preference of the negative $A_w$ branch.

Without technical priors and CMB information, the marginal posterior peaks around $A_w\simeq-0.01$ and $\alpha\simeq4$. The apparent preference for an intermediate value of $\alpha$ is, however, a prior-volume effect: the profile likelihood shows no corresponding maximum, but instead continues to improve as $\alpha$ approaches its limiting value, $\alpha\rightarrow3(4-\sqrt{6})$. Thus, after maximisation over $A_w$ and the remaining parameters, the data favours increasingly large $\alpha$, while the peak in the marginal posterior at $\alpha\simeq4$ arises from projection. 


With CMB-density priors and free $h$ (no $\theta_\star$ prior), the best-fit of $A_w$ is consistent with zero, hence the $\alpha$-profile is flat and its posterior contour is unconstrained. Once we impose the $\theta_*$-prior ($h$ is no longer free) the preference for the phantom behaviour is recovered, $\theta_*$ acts as a high-redshift lever arm that makes $\alpha$ identifiable again. In this case both the marginalised posterior and its profile likelihood peak at values further away from the $\Lambda$CDM limit, $A_w \sim -0.003$. Without CMB-information the matter density is slightly higher $\Omega_{\rm m}=0.3126^{+0.0096}_{-0.011}$ but consistent with $\Omega_{\rm m}=0.3003\pm 0.0038$ once CMB information is included.

There is no dependence on $z_q$ in the time-evolution of $w(z)$, only in its amplitude, where it is fully degenerate with $B$. Due to this degeneracy we cannot learn anything about this quantum-gravity parameter in power-law models. When introducing technical priors, the amplitude becomes strongly compressed as a function of $\alpha$ (see \cref{eq:alpha_of_m},  \cref{eq:msafe_prior} and \cref{app:parametr}): 
\begin{equation}
    A_{w, \rm max}(\alpha) =  M^2_{\rm safe} \frac{\alpha}{3 (1+z_{\rm in})^\alpha} m^2(\alpha) y_1(\alpha) \, . 
\end{equation}
As a result, the posterior in the equation-of-state parameters is mostly defined by this prior as shown in the left panel of \cref{fig:pl_model_contours_priors}, bringing the cosmological constraints to be driven towards $\Lambda$CDM. If we marginalise this posterior distribution, due to projections we would see a peak in $\alpha \sim 2$ $(m \sim -0.74)$. The apparent peak in this quantum-gravity parameter is a consequence of the technical prior ($m \rightarrow 0$ as $\alpha \rightarrow 0$ and $y_1 \rightarrow 0$ as $\alpha \rightarrow 4.65$), not a detection. Minimisation for the best-fit values reflects this by optimising $\alpha$ at the $A_w$-value maximally allowed by the technical prior. 

\begin{figure*}
    \centering
    \begin{minipage}[c]{0.4\textwidth}
        \centering
        \includegraphics[width=0.95\linewidth]
        {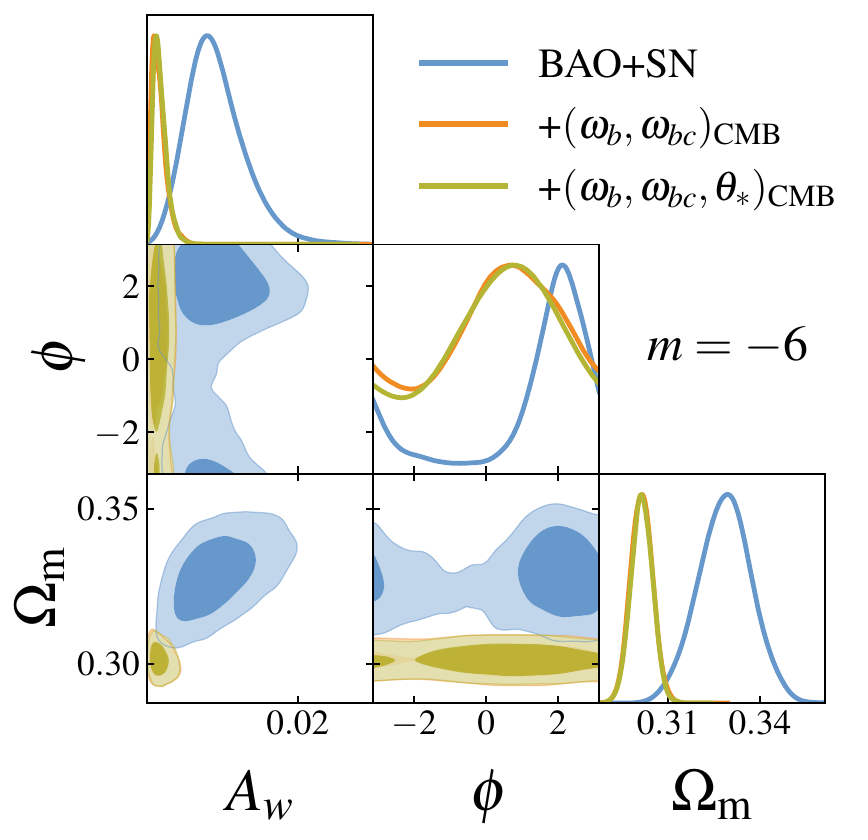}
    \end{minipage}
    \hfill
    \begin{minipage}[c]{0.59\textwidth}
        \centering
        \includegraphics[width=\linewidth]
        {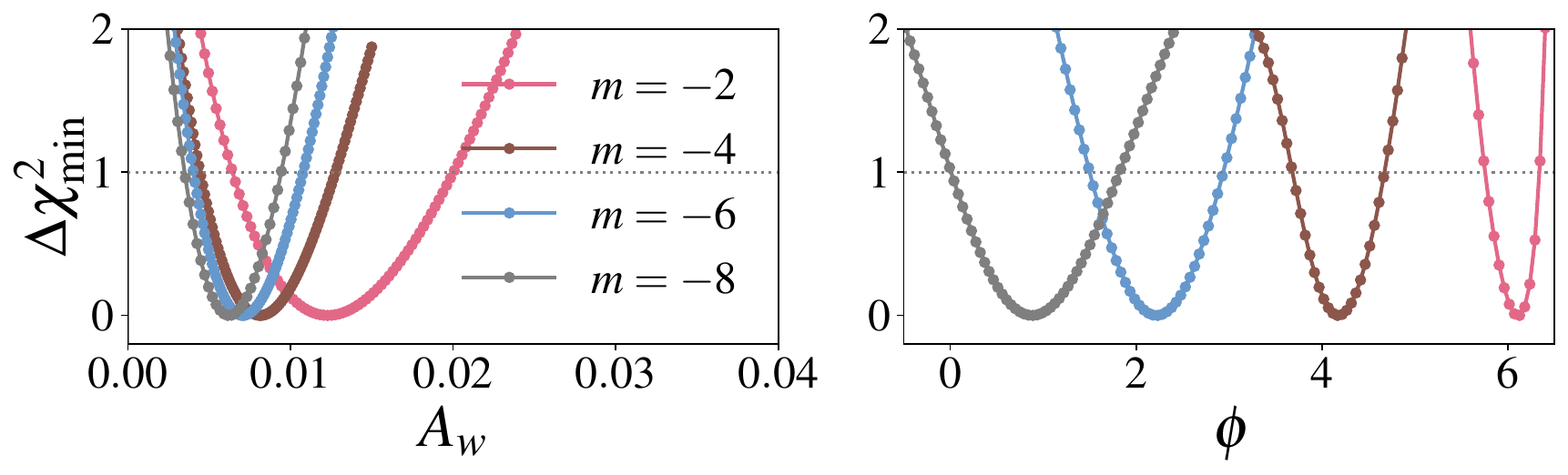}

        \vspace{0.5em}

        \includegraphics[width=\linewidth]
        {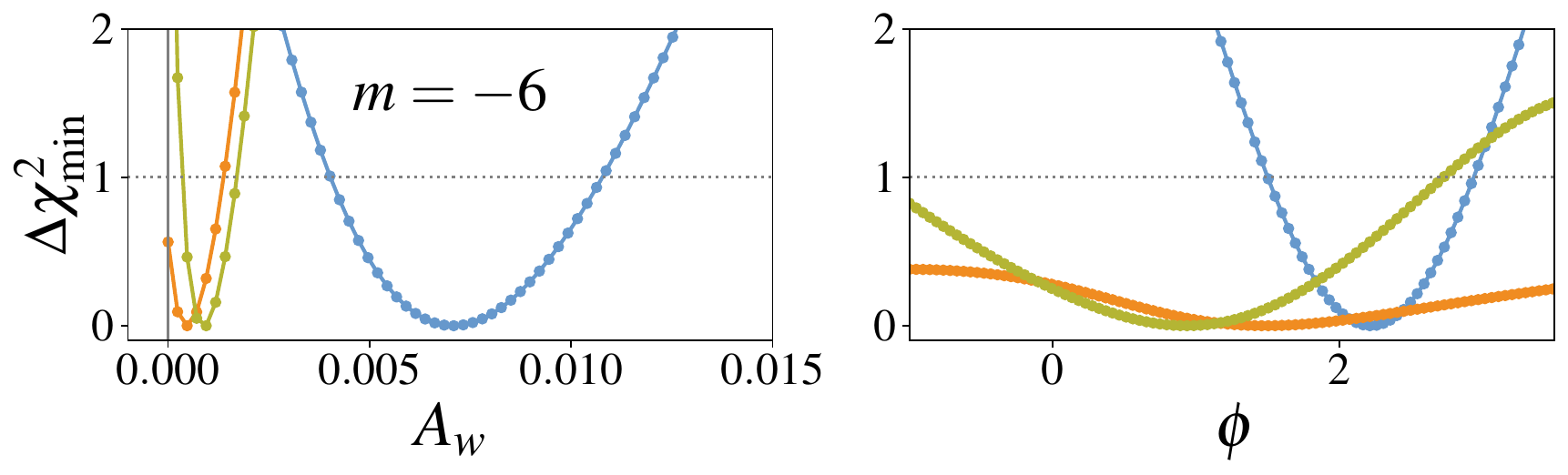}
    \end{minipage}

    \caption{Oscillatory model results with fixed frequencies. {\bf Left panel:} marginalised constraints on the dark-energy amplitude $A_w$ and phase $\phi$ for the three combinations of data specified in the legend in the quantum geometric case, $m=-6$. {\bf Lower right panel:} posterior profiling  for the $m=-6$ model in three datasets with the same colour-coding as in the left panel. {\bf Upper right panel:} posterior profiling  with SN+BAO without CMB information for the special cases given in the legend. All results are without technical priors.}
    \label{fig:osc_m_minus6_contours}
\end{figure*}

{\bf \textit{Oscillatory models}} exhibit the most interesting behaviour and provide the best fit for the datasets considered in this analysis among all three quantum-gravity models. We first consider fixed frequencies, i.e. fixed $m$ or $\beta$. In these cases, for the combination of BAO+SN without CMB-information, we obtain significant detections for some of the equation-of-state parameters. This is shown in the posterior-profiles (the upper right panel of \cref{fig:osc_m_minus6_contours}) for the special cases: $m=-2, -4, -6, -8$. 

Similar to the power-law models, their bets-fit values capture the dip in the DESI measurements at intermediate redshifts $z \sim 0.5-1$, similar to CPL (see \cref{fig:wz_models}). These models prefer high $\Omega_{\rm m} \sim 0.32$, as shown in the posterior distribution for the quantum geometric case $m=-6$ in the left panel of \cref{fig:osc_m_minus6_contours}. For BAO+SN data only, and for the oscillatory model $m=-6$, we obtain a better fit than CPL in the same setup. Once CMB information is added, $\Omega_{\rm m}$ and $h$ are anchored, shifting the peak in $A_w$ from $\sim 0.007$ to $\sim 10^{-3}$, since the CMB-density priors prevent  strong oscillations, and thus give poorer fits for higher-$z$ DESI distances. With these priors, and with the amplitude now tending to zero, the sensitivity to the phase $\phi$ decreases, as demonstrated in the posterior-profiles (the bottom right panel of \cref{fig:osc_m_minus6_contours}). In the full combination of the datasets, we have clear detections of equation-of-state parameters with a fit better than $\Lambda$CDM but worse than CPL for $m=-6$ without the technical priors.

When imposing the technical priors and varying $\log_{10}z_q$, these conditions translate to the $m$ or frequency-dependent upper bound on the amplitude:
\begin{equation}
\label{eq:a_w_max_osc}
    A_{w, \rm max}=\frac{(M_{\rm safe})^2 |\mathcal{M}(m)|}{(1+z_{\rm in})^6 f^2}\, ,
\end{equation}
where
\begin{align}
    f&=\max\left(|\cos\theta|,|\sin\theta|\right)\, ,\nonumber \\ 
    \theta(A,B) &= \operatorname{atan2}(B,A) = \arg(A+iB) \, , 
\end{align}
and $\mathcal{M}$ is given by \cref{eqn:mathcalm}. For $m=-6$, this gives $A_{w, \rm max} \approx 8.8 \times 10^{-5}$ for all phases and $1.76 \times 10^{-4}$ for diagonal directions in $(A,B)$.

The right panel in \cref{fig:pl_model_contours_priors} illustrates the impact of the frequency-dependent technical prior on the oscillatory model. The dashed lines indicate the limiting amplitude bounds imposed by the technical prior, which depends on the oscillation frequency, $m$ or $\beta(m)$. The resulting posterior retains a non-trivial structure within the allowed region, and is therefore not completely prior dominated. Small fractions of the $2\sigma$ region without the technical prior remain allowed (e.g. the probability mass of $A_w$ lower than $A_{w, \rm max}$, with CMB density priors, is $\sim 10\%$ and with the full-compressed CMB information $\sim 3\%$). The best-fit amplitude in this case is the maximum bound of the amplitude and the phase is optimised correspondingly.


\begin{figure*}
    \centering
    \begin{minipage}[c]{0.5\textwidth}
        \centering
        \includegraphics[width=\linewidth]
        {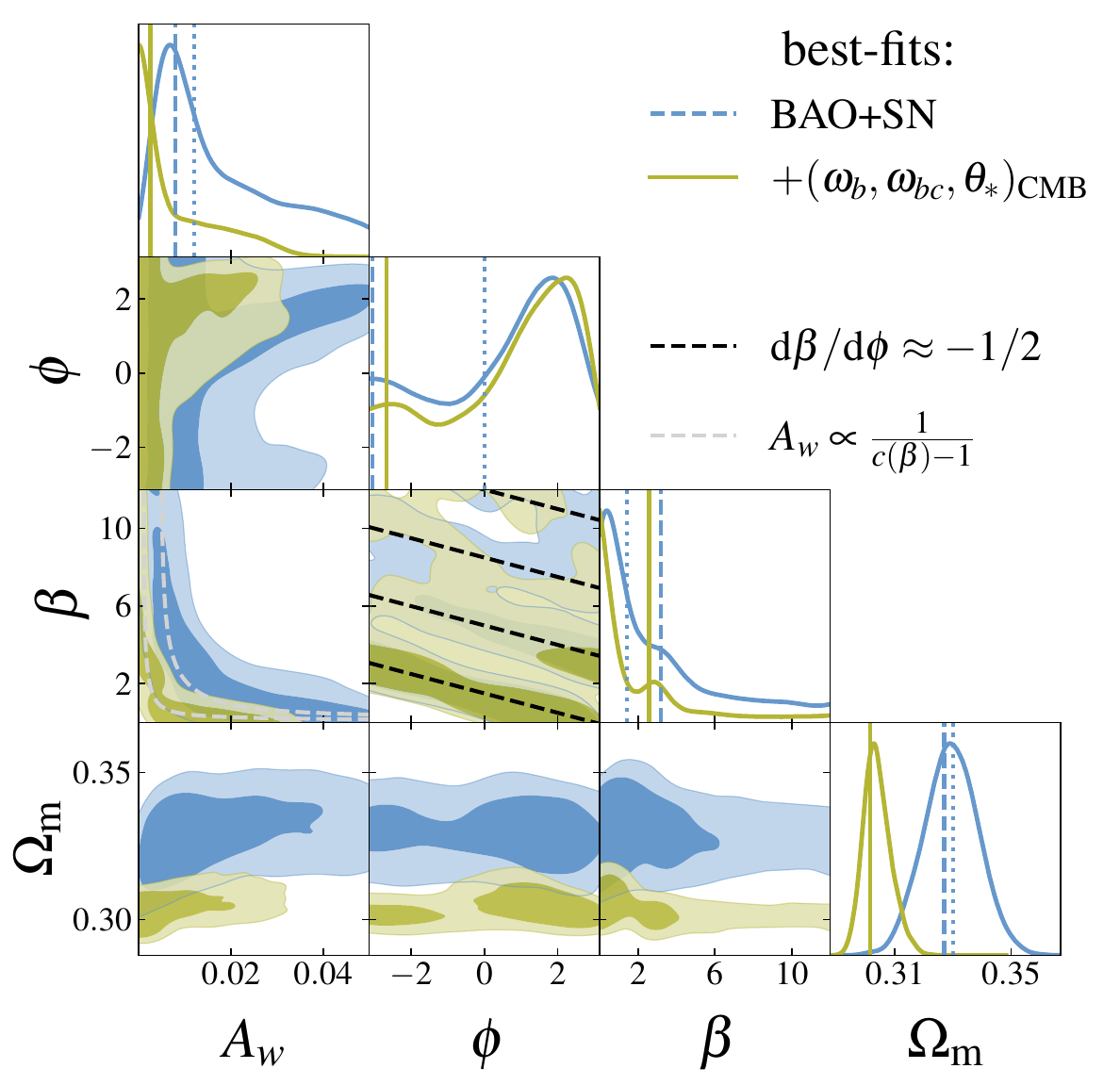}
    \end{minipage}
    \hfill
    \begin{minipage}[c]{0.49\textwidth}
        \centering
        \includegraphics[width=0.5\linewidth]
        {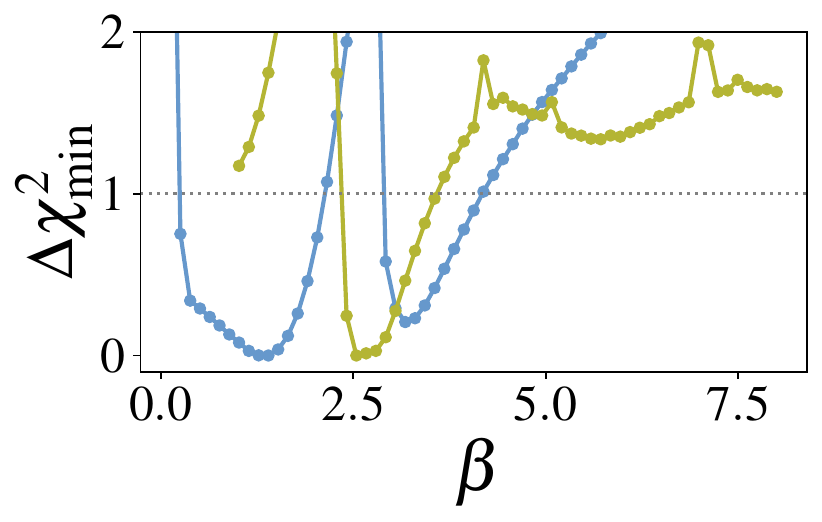}
        \vspace{0.1em}

    \includegraphics[width=0.9\linewidth]
        {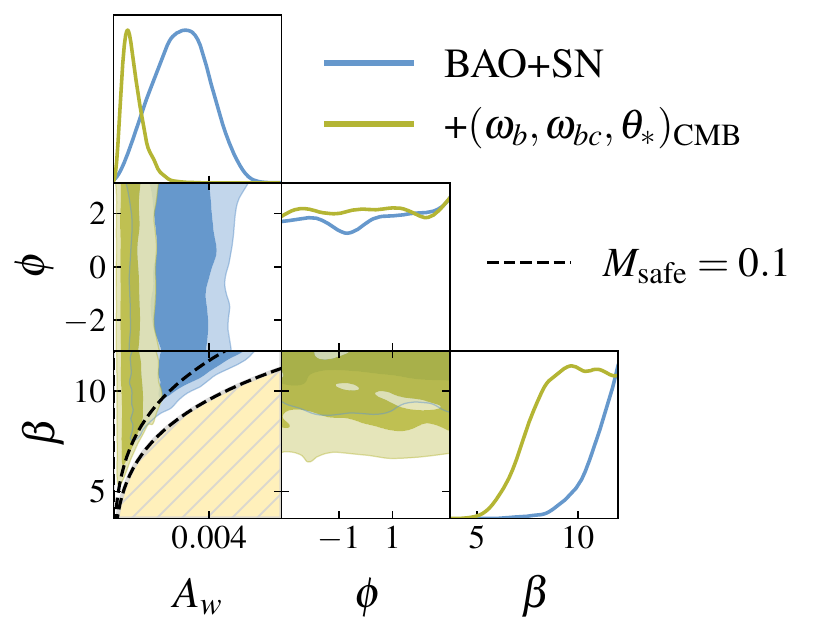}
    \end{minipage}

   \caption{Oscillatory model results with varying frequency. {\bf Left panel:} marginalised constraints on the dark-energy amplitude $A_w$, frequency $\beta$, phase $\phi$ and matter density $\Omega_{\rm m}$ for the two combinations of data specified in the legend. Vertical dashed, dotted and solid lines represent the best-fit values, while black and grey dashed lines denote degeneracies discussed in the main text.  {\bf Upper right panel:} posterior profiling  in frequency for the same datasets as in the left panel. Both results are without technical priors. {\bf Lower right panel:} marginalised constraints on the equation-of-state parameters with the technical prior on $(A,B)$ with varying $\log_{10}z_q$. The black dashed lines correspond to $M_{\rm safe}=0.1$ for $f=1$ and $f=1/\sqrt{2}$ from \cref{eq:a_w_max_osc}.}
    \label{fig:osc_model_contours}
\end{figure*}

Next we open the parameter space and vary the frequency parameter, $\beta$.
The smooth dip seen in BAO+SN data can be realised through different combinations of $(A_w, \phi, \beta)$, see \cref{fig:wz_models}. For instance, the minimiser finds two preferred points, $\beta \sim 1.4$ $(m\sim -2)$ and $\beta \sim 3.2$ $(m \sim -5)$ (see upper right panel of \cref{fig:osc_model_contours}). In other words, the same phantom-dip behaviour is equally well reproduced with two different frequencies and their corresponding set of amplitudes and phases (dashed and dotted blue lines in the left panel of \cref{fig:osc_model_contours}). 

As we have seen in the special case of $m=-6$, adding CMB density priors, keeping $h$ free, pulls the amplitude towards zero and decreases sensitivity to the phase, as well as to the frequency. Also this combination pins down $\Omega_{\rm m}h^2$, which is similar for both $m\sim -2$ and $m \sim -5$ as we know from our minimisation runs with fixed $m$. To resolve the oscillation's frequency, we must impose the CMB distance $\theta_*$ prior. In this case we have only one clear maximum $\beta \sim 2.5 \, (m\sim -3.5)$ with a plateau at $\beta \geq 4$. 

In the posterior distribution (left panel of \cref{fig:osc_model_contours}), we highlight interesting features in the oscillating models, which capture the $w(z)$ phantom-dip behaviour preferred by the data. First, the data can actually pin down the argument of the cosine (\cref{eqn:oscnewpara})
\begin{equation}
    \psi(z) \equiv 3\beta \log{(1+z)}+\phi \approx \text{const.} \, , 
\end{equation}
within the redshift range probed by DESI, so that $\psi(z)$ produces an anti-correlation between the phase and the frequency (constant effective phase) at a fixed redshift, $z_{\rm eff}$
\begin{equation}
    \phi \approx \phi_0 - 3\beta \log{(1+z_{\rm eff})} \, .
\end{equation}
Since $\phi$ is plotted in one $2\pi$ interval, this anti-correlation line appears multiple times. For $z_{\rm eff} \sim 1$ with $\beta \in [0, 12]$, we get four lines (see \cref{fig:osc_model_contours}). For $z_{\rm eff} = 1$,  the slope in the anti-correlation lines is given by $\mathrm{d}\beta/\mathrm{d} \phi \sim -1/2$, which is the slope chosen for the lines in posteriors in \cref{fig:osc_model_contours}. These lines effectively represent the same $w(z)$ shape in the data window.

Of course $c(\beta)$ (\cref{eqn:oscnewpara})  varies with $\beta$ too. This can be clearly observed in the 2D $A_w-\beta$ plane (constant phantom-dip): 
\begin{equation}
    A_w (1+z_{\rm eff})^6 \left[ c(\beta) +\cos{\psi (z_{\rm eff})} \right] = \delta w(z_{\rm eff}) \approx \text{const.}\,,
\end{equation}
with $m=-(4+\beta^2)/3$. As discussed in \cref{sec:gftmodels_reparam}, one can see that $c(\beta)$ falls from $1$ at $\beta \rightarrow 0$ quickly to $-1$ at $\beta \sim 3.74$ and then converges towards $-2$ when $\beta \rightarrow \infty$. By choosing $\phi(\beta)$ so that the cosine is at a minimum near $z_{\rm eff}$, we can recreate the shape observed in the contours:
\begin{equation}
    A_w(\beta) \sim \frac{\delta w(z_{\rm eff})}{(1+z_{\rm eff})^6 (c(\beta)-1)},
\end{equation}
which produces a large $A_w$ as $\beta \rightarrow 0$, giving a low-frequency, high-amplitude family of models. Similarly,  when $A_w\sim -\delta w(1)/192$ we have $\beta \rightarrow \infty$ which is a high-frequency, low-amplitude family of models. This is represented by the dashed and dotted blue lines in \cref{fig:osc_model_contours}. Overall, these complex degeneracies introduce projection effects, which we see via discrepancies between the best-fits and the marginalised peaks.

Once we include the technical priors the picture changes: the particular structures in the $\phi-\beta$ plane and $A_w-\beta$ plane  disappear because $M_{\rm safe}=0.1$ forbids the $(A_w, \beta)$ region which can produce it (see right-hand side of \cref{fig:osc_model_contours}). The maximum amplitude is now a function of $\beta$ as per \cref{eq:a_w_max_osc}, and the previous best-fits are excluded by the prior. The other consequence of the technical prior is that high $\beta$ (more negative $m$) values are preferred, as this allows for a non-$\Lambda$CDM limit in the amplitude. 

For BAO+SN, the peak in the marginalised amplitude $A_w$ is a prior-volume effect due to the accumulation of volume at the $\beta$ prior-maximum. With CMB information, the marginalised peak close to the $\Lambda$CDM limit in $A_w$ is a consequence of sampling in $X=A_w \cos{\phi}$ and $Y=A_w \sin{\phi}$, i.e. the ``informative prior'' from \cref{eq:a_w_from_x_y}. 

When we impose technical priors (right hand side of \cref{fig:osc_model_contours}), we have $A_{w} \rightarrow A_{w,\rm max}(\beta)$ not $A_w \propto 1/(c(\beta)-1)$. Note that the original posterior form can be recovered if $M_{\rm safe}$ is of order one. If we were to sample $A_w$ uniformly instead of $(X,Y)$, both 1D posteriors would peak at zero. The CMB-informed likelihood already dislikes high-frequency models, so increasing $\beta$ above 12 will just make the volume artefacts worse. Similar to all quantum-gravity models, $z_q$ is unconstrained again. However, it does control the time-evolution in $w(z)$, so that with more information one could put constraints on $z_q$ if the phase is constrained.

\begin{table*}
\centering
\caption{
Selected marginal constraints (weighted means and 68 per cent unless an inequality is shown)
and best-fit values for the analyses considered in this work: BAO+SN+$(\omega_{\rm b}, \omega_{\rm bc}, \theta_*)_{\rm CMB}$.
The best-fit values correspond to the maximum-likelihood points and
need not coincide with the marginal posterior constraints due to
projection effects. The \(A,B\) prior refers to the
technical prior given in
\cref{eq:msafe_prior}. The difference in best-fit $\chi^2$ is evaluated at the maximum a posteriori (MAP) point relative to $\Lambda$CDM, $\Delta\chi^2_{\min}=\chi^2_{\min}-\chi^2_{\min,\Lambda{\rm CDM}}$, such that negative values indicate a better fit than $\Lambda$CDM. Note that the quantum-gravity parameter $z_q$ is unconstrained when varied in all models, hence the effective number of dark-energy parameters is $N_{\rm par}-1$.
}
\label{tab:constraints}
\begin{tabularx}{\linewidth}{lccllc}
\toprule
Model & $N_{\rm par}$ & $(\Omega_{\rm m},H_0)$
& Marginal QG/DE constraints
& Best-fit QG/DE values
& $\Delta \chi^2_{\min}$ \\[.5mm]
\hline
\midrule
$\Lambda$CDM
  & - & $(0.3025\pm 0.0035, 68.18\pm 0.26)$ & -- & -- & 0. \\
\midrule

CPL
  & 2 & $(0.3114\pm 0.0056, 67.48^{+0.54}_{-0.61})$ & $(w_0,w_a):\ -0.846\pm 0.055, -0.56\pm 0.21$
  & $(-0.851, -0.54)$ & -5.6 \\
\midrule
Log
  & 1 & $(0.3025\pm 0.0035, 68.11\pm 0.27)$ & $\widetilde A_w < 7.67\cdot 10^{-5}$
  & $2\times10^{-13}$ & 0.2 \\

  with $B$ prior
  & 2 &  &  $(\widetilde A_w, \log_{10}z_q)$
  & $(3\times10^{-10}, 0.482)$ & 0.2 \\
\midrule
Power law
  & 2 & $(0.3003\pm 0.0038, 68.71^{+0.38}_{-0.45})$ &
  $(A_w,\alpha):\ -0.0083^{+0.0081}_{-0.0014}, > 2.06| 4.65$
  & $(-0.003, 4.65)$ & -0.8 \\

  with $B$ prior
  & 3 &  & $(A_w,\alpha, \log_{10}z_q)$
  & $(-0.00001, 3.44, 0.906)$ & 0.2 \\

\midrule
Oscillatory, $m=-6$
  & 2 & $(0.3014\pm 0.0036, 68.91^{+0.33}_{-0.38})$ &
  $(A_w,\phi):\ (0.0017^{+0.0005}_{-0.0011}, 0.3^{+2.1}_{-1.2})$
  & $(0.0009, 0.933)$ & -1.9 \\

  with $A,B$ prior
  & 3 &  & $(A_w,\phi, \log_{10}z_q)$
  & $(0.0002, 0.349, 0.800)$ & -1.0 \\
\midrule
Oscillatory
  & 3 & $(0.3040^{+0.0038}_{-0.0054}, 68.62\pm 0.46)$ &
  $(A_w,\beta,\phi):\ < 0.0116,< 2.72,0.62^{+2.5}_{-0.67}$
  & $(0.0025,2.575,-2.662)$ & -3.0 \\

  with $A,B$ prior
  & 4 &  &$(A_w,\beta,\phi, \log_{10}z_q)$
  & $(0.0005, 9.047, 0, 0.808)$ & -1.6 \\

\bottomrule
\end{tabularx}
\end{table*}

\section{Conclusions}
\label{sec:conclusions}


We have presented the first direct observational constraints on the
late-time dark-energy dynamics emerging from  GFT quantum
gravity, using DESI DR2 BAO and Pantheon+ supernovae, with and without
compressed CMB information. The three branches of the theory lead to
markedly different phenomenology. The logarithmic branch, which permits
only $w>-1$, is driven extremely close to the $\Lambda$CDM limit, while
the power-law branch can accommodate the mild preference of the data for
phantom evolution. The oscillatory branch provides the best fits among
the GFT models considered here: without  technical perturbative
priors, its additional freedom allows it to reproduce the intermediate-redshift
phantom dip favoured by the distance data. When the oscillation frequency
is varied, BAO+SN admit preferred solutions around $m\sim-2$ and
$m\sim-5$, while the inclusion of CMB information selects a broader
preference around $m\sim-3.5$, where $m$ is the GFT microscopic interaction parameter. Very conservative technical priors enforcing
the perturbative regime substantially restrict the allowed amplitudes
and drive all three branches closer to $\Lambda$CDM.

The value $m=-6$, which is particularly well motivated from the perspective of discrete-gravity path integrals, therefore does not coincide with either best fit, although the preference without CMB priors lies relatively close to it. Crucially, both best-fit values remain within the same oscillatory regime as $m=-6$ and thus produce the same qualitative form of the dark-energy dynamics. Quantum-geometric models ($m=-6$) therefore provide a compelling qualitative template for emergent dynamical dark energy, emphasising the importance of their theoretical development. At the same time, the preference for somewhat larger values of $m$ motivates investigating which mechanisms within GFT could shift the effective interaction parameter away from $m=-6$ in the direction favoured by the data. Since GFT interactions encode the atoms of the cellular complexes entering discrete-gravity path integrals \citep{Oriti:2014yla}, it is equally important to identify the combinatorial counterpart of such mechanisms and the corresponding generalisation of the underlying discrete structures. Our results therefore point to concrete directions for further research not only in GFT, but also in simplicial quantum gravity and spin-foam model building.

An important lesson of this analysis is that marginal posterior
distributions alone can give a misleading picture of the information
contained in the data. The highly non-Gaussian parameter degeneracies,
together with the non-trivial mapping between the microscopic GFT and
observational parameters, generate substantial prior-volume and
projection effects. Apparent posterior preferences for particular
values of the GFT parameters can therefore occur even when the
corresponding profile likelihood is essentially flat. In particular,
$z_q$, the parameter controlling the average number of quantum gravity atoms, remains unconstrained by the present data: much of its dependence
is absorbed into the observable amplitude and, for the oscillatory
branch, the phase. Constraining this genuinely quantum-gravitational
parameter will require a significant detection of non-$\Lambda$CDM
evolution together with greater leverage on its redshift dependence. 
Thus, it will 
benefit from extending the
redshift lever arm in both directions: improved low-redshift measurements, including peculiar velocities and future radio surveys can better resolve the late-time evolution, while higher-redshift probes such as Ly$\alpha$ measurements can provide the longer baseline needed to distinguish its characteristic redshift dependence.

The preference for non-monotonic and phantom-crossing behaviour found
here is qualitatively consistent with a broader literature on dynamical
dark energy. Model-independent reconstructions have long found that
current distance data can accommodate evolving and phantom-crossing
equations of state~\citep{Zhao:2017cud}, while recent analyses including DESI have recovered similar behaviour using non-parametric
reconstructions of either $w(z)$ or the dark-energy
density~\citep{Ormondroyd:2025iaf,Berti:2025phi}. Oscillatory
parametrisations have likewise been found to improve the fit relative
to $\Lambda$CDM, with recent DESI DR2 analyses finding particular
support for simple models exhibiting late-time oscillatory
features~\citep{Escamilla:2024fzq,Kessler:2025kju}. Our results place
these phenomenological indications in a different context: the
non-monotonic evolution and phantom crossing arise here as predictions
of an underlying quantum-gravity construction, allowing the same
observations to constrain parameters with a direct microscopic
interpretation.

Two extensions are particularly important. First, greater analytic control over the non-perturbative mean-field GFT dynamics would be valuable. Although numerical analyses indicate that the non-perturbative solutions closely track the perturbative ones \citep{Marchetti:2025jze}, an analytic understanding would allow their domain of validity to be established more robustly and the resulting theoretical priors to be determined more accurately. Second, the analysis should be extended beyond the homogeneous sector. An immediate phenomenological step would be to retain the minimal early-time completion adopted here, in which the background reduces to $\Lambda$CDM, while evolving perturbations according to the standard GR equations on a background modified by quantum-gravity effects only at late times. Closely related hybrid prescriptions are commonly employed in effective approaches to quantum-gravity cosmology, notably in loop quantum cosmology~\citep{Agullo:2023rqq}. Ultimately, however, the perturbation dynamics should be derived from the underlying GFT theory. Even if direct quantum-gravity corrections to perturbations prove negligible at low energies, establishing this robustly would justify the use of perturbation-sensitive datasets and thereby substantially sharpen the constraints obtained here.

In conclusion, this work represents an important advance in connecting non-perturbative quantum gravity with low-energy cosmological observations. In doing so, it takes a first step towards establishing a systematic and constructive feedback loop between observations and quantum-gravity model building.

\section*{Acknowledgements}

 BB is supported by a UK Research and Innovation Stephen Hawking Fellowship (EP/W005654/2). MT's research is supported by grant ST/Y000986/1. L.M.~acknowledges support from the Kavli Institute for the Physics and Mathematics of the Universe, from the Okinawa Institute of Science and Technology Graduate University, and from the John Templeton Foundation, through ID\# 62312 grant as part of the \href{https://www.templeton.org/grant/the-quantum-information-structure-of-spacetime-qiss-second-phase}{\textit{`The Quantum Information Structure of Spacetime'} Project (QISS)}. For the purpose of open access, the author has applied a Creative Commons Attribution (CC BY) license to any Author Accepted Manuscript version arising from this submission. 

\section*{Data Availability}

Links and references with sources of the data and analysis pipelines are provided in the main text. 



\bibliographystyle{mnras}
\bibliography{refs} 

\appendix

\section{parametrisations of emergent dark energy models}\label{app:parametr}
As discussed in the main text, parametrizing $\delta w(z)$ in terms of $\{m,z_q,A,B\}$, although natural from the perspective of the underlying quantum-gravity dynamics, is not particularly effective for comparing the resulting models with the background data considered in this work. In this appendix, we provide explicit relations between the natural parametrisation introduced in \citet{Marchetti:2025jze} and that adopted in \cref{sec:samp_param}.

\subsection{Physical parametrisation}
We begin by presenting the expressions of \cite{Marchetti:2025jze} for $\delta w(z)$ in terms of the natural  parameters $\{m,z_q, A,B\}$.
\paragraph*{\ref{item:osc} Oscillatory models ($1 + 3m/4 < 0$):}
\begin{align}
\delta w(z) = & \left(\frac{1+z_q}{1+z}\right)^{-6}
 \nonumber \\ & \times \left[
\delta w_0 + \delta w_1 \cos\big(2\Phi(z)\big) + \delta w_2 \sin\big(2\Phi(z)\big)
\right] \, ,
\label{eq:oscwz}
\end{align}
where the phase evolves logarithmically with redshift
\begin{equation}
\label{eq:Phi}
\Phi(z) = \frac{3 \beta}{2} \log\left(\frac{1+z_q}{1+z}\right) \, ,\qquad \beta \equiv 2 \sqrt{-1 -3m/4}\,.    
\end{equation}
The parameters $\delta w_i = \delta w_i (A,B;m)$ can be expressed in terms of $(A,B,m)$ by introducing first
\begin{subequations}
\begin{align}
c_A &\equiv A - \frac{\beta}{2}B \, , 
\qquad
&c_B &\equiv B + \frac{\beta}{2}A \, , \\
\kappa_A &\equiv \frac{3}{2}m^2 A^2 - 4c_A^2 \, , 
\qquad
&\kappa_B& \equiv \frac{3}{2}m^2 B^2 - 4c_B^2 \,, \\
\kappa_{AB} &\equiv 3m^2 AB - 8c_A c_B \,.
\end{align}
\end{subequations}
In terms of these quantities, $\delta w_i(A,B,m)$ read 
\begin{subequations}
\begin{align}
\delta w_0 &= -\left(\kappa_A+\kappa_B\right) \, ,  \\ 
\delta w_1& =
-\frac{(2-\beta^2)(\kappa_A-\kappa_B)-3\beta \kappa_{AB}}
{2(1+\beta^2)} \, ,   \\ 
\delta w_2 &=
-\frac{3\beta(\kappa_A-\kappa_B)+(2-\beta^2)\kappa_{AB}}
{2(1+\beta^2)} \, .
\end{align}
\end{subequations}

\paragraph*{\ref{item:pl} Power-law models ($-4/3<m\le 0$):}
\begin{equation}
\delta w(z) = \delta \tilde{w}_0 \left(\frac{1+z_q}{1+z}\right)^{-\alpha} \, ,
\label{eq:plwz}
\end{equation}
with $\alpha = 6(1-\mu)$, and $\mu \equiv \sqrt{1+3m/4}$, while $\delta \tilde{w}_0$ is given by 
\begin{equation}
    \delta \tilde{w}_0 = 2 m^2 B^2 (1-\mu) \, y_1(m) \, ,
    \label{eq:amplitude}
\end{equation}
where 
\begin{equation}
\label{eq:y1_m}
 y_1(m) = -\frac{2}{1+2\mu} \left[\frac{3}{8} - \frac{(1-\mu)^2}{m^2} \right]   \, .
\end{equation}
Note that $\delta\tilde{w}_0\ge 0$ for $m\le 2(1-2\sqrt{6}/3)$.
\paragraph*{\ref{item:log} Logarithmic models ($1 + 3m/4 = 0$):}
\begin{equation}
\delta w(z) = 9 \delta \tilde{w}_0 \left(\frac{1+z_q}{1+z}\right)^{-6}
\log^2\left(\frac{1+z_q}{1+z}\right) \, ,
\label{eq:logwz}
\end{equation}
where $\delta \tilde{w}_0$ is given by \cref{eq:amplitude}. Note that since in this case $\mu=0$ and $y_1 =3/8$, we have $\delta \tilde{w}_0>0$.

Let us remark that in general Power-law and Logarithmic models depend on both the initial conditions $A$ and $B$ for the two solutions of the perturbative mean-field quantum gravity dynamics \citep{Marchetti:2025jze}. However, here we are assuming that the growing solution dominates, which is reasonable at late times, effectively reducing the dependence of Power-law and Logarithmic models to $\{B,m,z_q\}$ only.

\subsection{Sampling parametrisations}
\label{sec:gftmodels_reparam}
Below we connect the above expressions with the parametrisation adopted in  \cref{sec:samp_param}, based on the decomposition of $\delta w(z)$ as a product of an amplitude and time dependent function:
\begin{equation}
     \delta w(z) =  A_wf_w(z)\,. 
\end{equation}
\paragraph*{\ref{item:osc} Oscillatory models:}
In this case, we perform a complex decomposition of the functional form. Starting from \cref{eq:oscwz}, one can rewrite
\begin{equation}
    \delta w_1\cos(2\Phi(z))+\delta w_2\sin (2\Phi(z))=r\cos(3\beta\log (1+z)+\phi)\,,
\end{equation}
with $r\ge 0$ and
\begin{equation}
    r e^{i\phi}e^{-3i\beta\log(1+z_q)}=(\delta w_1+i\delta w_2)=(A+i B)^2 \mathcal{M}(\beta)\,.
\end{equation}
Explicitly,
\begin{align}\label{eqn:mathcalm}
   \mathcal{M}(\beta)&\equiv \frac{(\beta -2 i)^3 (2+\beta  (\beta +4 i)) }{12 (\beta +i)}\,.
\end{align}
Moreover, $\delta w_0$ can be expanded as
\begin{equation}
    \delta w_0=-\frac{(A^2+B^2)}{6}(\beta^4-2\beta^2-8)\,.
\end{equation}
This allows us to recast \cref{eq:oscwz} in the form of \cref{eqn:oscnewpara} by identifying
\begin{subequations}
\label{eqn:parametersoscnew}
\begin{align}
    A_w(A,B,\beta,z_q)&= r(A,B,\beta) (1+z_q)^{-6}\,,\\
    c(\beta)&=-\frac{(\beta^4-2\beta^2-8)}{6\vert \mathcal{M}(\beta)\vert}\,,\label{eqn:cofbeta}\\
    \phi(A,B,\beta)&=2\arg(A+iB)+\arg\mathcal{M}(\beta)-3\beta\log(1+z_q)\,,
\end{align}
\end{subequations}
with $A_w\ge 0$. As $c(\beta)$ is a function of $\beta$ only, it is clear that $\delta w(z)$ can be described by the three independent parameters $\{A_w,\beta,\phi\}$. Moreover, $c(\beta)$\footnote{Additionally, $c(\beta)\rightarrow -2$ for $m\rightarrow -\infty$.} is a monotonic function that takes values between $-1$ and $1$ for $-6\le m<-4/3$, showing that for each $m$ in this range there is a value of $\phi$ which leads to $f_w(z_{\rm piv})=0$. If we choose to sample in $w_{\rm piv}$, this will lead to amplitudes $A_w \propto f_w(z_{\rm piv})^{-1} \rightarrow \infty$, which can yield a very poor fit. This highlights why we choose \emph{not} to sample in $w_{\rm piv}$. It is still good as a derived parameter, since the likelihood is sensitive to $w(z)$ around $z \sim 0.5-1$.

\paragraph*{\ref{item:pl} Power-law models:} For Power-law models, comparing \cref{eqn:timedeppl} and \cref{eq:plwz} immediately allows us to identify 
\begin{equation}
\label{eq:pl_amp}
A_w =
\epsilon_\alpha \frac{\vert \delta \tilde{w}_0\vert}{(1+z_q)^{\alpha}},
\end{equation}
where $\epsilon_\alpha = {\rm sgn}[\alpha - 3(4 - \sqrt{6})]$ and $\alpha (m) = 6 ( 1- \sqrt{1+3m/4})$.
\paragraph*{\ref{item:log} Logarithmic models:} Finally, comparing \cref{eqn:timedeplog} and \eqref{eq:logwz}, we obtain for the Logarithmic models
\begin{equation}
    A_w=\frac{12B^2}{(1+z_q)^6}\ge 0\,.
\end{equation}
In this case, the amplitude is set by \(B\) and \(z_q\), with the latter also controlling the functional form of the redshift evolution.

\end{document}